\documentclass[12pt,a4paper]{article}
\usepackage[T2A]{fontenc}
\usepackage[english]{babel}
\usepackage{graphicx}
\usepackage{float}
\usepackage[a4paper,top=3cm,bottom=2cm,left=1.5cm,right=1cm,marginparwidth=1.75cm]{geometry}
\usepackage[colorlinks]{hyperref}

\usepackage{amsmath}
\usepackage{amssymb}
\usepackage{authblk}

\title{Heteroclinic cycles and chaos in a system of four identical phase oscillators with global biharmonic coupling}
\author[1]{\small Aleksei M. Arefev}
\author[1]{\small Evgeny A. Grines}
\author[1]{\small Grigory V. Osipov}
\affil[1]{\footnotesize Lobachevsky   State   University   of   Nizhni   Novgorod,   23
Gagarin av., Nizhny Novgorod 603950, Russia}
\affil[ ]{\textit{evgenij.grines@gmail.com}}
\date{}

\begin{document}
\maketitle

\begin{abstract}
We study a system of four identical globally coupled phase oscillators with biharmonic coupling function. 
Its dimension and the type of coupling make it the minimal system of Kuramoto-type (both in the sense of the phase space's dimension and the number of harmonics)  that supports chaotic dynamics.
However, to the best of our knowledge, there is still no numerical evidence for the existence of
chaos in this system.
The dynamics of such systems is tightly connected with the action of the symmetry group on its phase space.
The presence of symmetries might lead to an emergence of chaos due to scenarios involving specific heteroclinic cycles.
We suggest an approach for searching such heteroclinic cycles and showcase first examples of chaos in this system found by using this approach.
\end{abstract}


\section*{Introduction}
Systems of interacting oscillators are a classical and relevant object of study in nonlinear dynamics to this day.
The study of the behavior of systems of coupled phase oscillators occupies a special place in this research field.
A phase model was first introduced by Winfree \cite{Winfree1967, winfree1980} as a phenomenological explanation \cite{ BerYanMaiSch2021,LeonPazo2022} for an emergence of synchronisation in populations of weakly coupled limit cycle oscillators.
Subsequently, the model proposed by Kuramoto opened the way for a large number of both analytical and numerical studies of various phenomena of collective dynamics \cite{AceBonVicRitSpig2005, PikRos2015}.
A natural generalization of this model are Kuramoto-type systems
\begin{equation}
\label{eq:KurTypeGeneral}
\dot{\varphi}_i = \omega_i + \sum_{j = 1}^{N} A_{ij}\, g_{ij}(\varphi_i - \varphi_j),
\end{equation}
whose dynamics is of considerable interest to researchers.

In Kuramoto-type systems the complexity and diversity of the resulting dynamics depends on the heterogeneities (either in natural frequencies $\omega_i$ of phase oscillators or in interactions between them), the ensemble topology (which is encoded by constants $A_{ij}$) and the coupling functions $g_{ij}(\phi)$ themselves \cite{StankPerMcClintStefa2017}.
The simplest coupling with a single first harmonic (Kuramoto coupling \cite{Kur1975, kuramoto1984chemical} $g_{ij}(\phi) = -\sin{\phi}$ or Kuramoto-Sakaguchi coupling \cite{KurSak1986} $g_{ij}(\phi) = -\sin{(\phi - \alpha)}$) already can lead to  non-trivial dynamics when natural frequencies are non-identical or an ensemble has non-trivial topology. 
Quasi-periodic, chaotic and hyperchaotic dynamics can emerge in the case of globally coupled oscillators with strong detuning of natural frequencies \cite{PopMaistrTass2005}.
Chaotic dynamics can also be found in a chain of phase oscillators with uniform detuning studied by Topaj and Pikovsky \cite{TopPik2002}.
If all oscillators have the same natural frequency, the same Kuramoto-Sakaguchi coupling can lead to complex dynamics due to heterogeneities in the interactions of oscillators or a non-trivial topology of connections \cite{Burylko2021}.
In \cite{BiPaMa2018, BurMarBick2022} chaotic dynamics is observed in a system divided into several populations: different coupling functions are responsible for the interactions of oscillators within and between populations.
The dynamics observed in these works can be interpreted as a weak chimera \cite{AshBur2015, BickAsh2016, Bick2017} -- trajectories on their attractors exhibit localized frequency synchrony.
Although the primary context for our work is Kuramoto-type systems \eqref{eq:KurTypeGeneral}, which include only pairwise interactions between oscillators, it is important to mention the modern line of research that takes into account non-pairwise interactions (see the reviews \cite{BickGroHarScha2021, MajPercGhosh2022}).
Such systems naturally arise both in the formulation of certain problems and as the phase reduction of ensembles of interacting oscillators \cite{AshRodr2016,  LeonPazo2019,GenTeiRosPik2020}.
In addition to chaotic dynamics \cite{BAR16}, such systems can also exhibit structurally stable attractive heteroclinic cycles between weak chimeras \cite{Bick2018, Bick2019, BickLohse2019}.
If both the natural frequencies of the oscillators and the interactions between them are identical, then the non-local coupling between the phase oscillators can also lead \cite{WolfOmel2011, SudaOkuda2015} to the emergence of chimera states \cite{Parastesh2021, KurBat2002, AbrStrog2004,Omel2018,Haug2021}.

In the most symmetric case (globally coupled identical phase oscillators -- all oscillators are coupled to each other with the same coupling function and have the same natural frequency)
\begin{equation}
\label{eq:KurTypeGlobalIdentical}
\dot{\varphi}_i = \omega + \sum_{j = 1}^{N} g(\varphi_i - \varphi_j),
\end{equation}
the coupling function becomes the only source of the complexity and diversity of the dynamics. 
In that case Kuramoto and Kuramoto-Sakaguchi couplings lead \cite{BerYanMaiSch2021} to a partially-integrable system \cite{WatStro1993}, whose dynamics is essentially two-dimensional \cite{BurMarBick2022} regardless of the number of oscillators. Later it was shown that only certain types of equilibria or limit cycles could be attractors in this system for $N > 3$ \cite{EngelMir2014}.
However, using the biharmonic coupling (that is, including the second harmonic into the coupling function) introduces a variety \cite{CPR16} into the possible dynamics of the system, leading, for example, to the appearance of structurally stable heteroclinic cycles as attractors \cite{HMM93, KK01}
The example of chaotic dynamics in systems of globally coupled identical phase oscillators with the biharmonic coupling function was first demonstrated in \cite{AOWT07}.
This chaos is tightly interconnected with this system being reversible \cite{ABB16, burylko2020collective} at certain parameter values, which allows to conclude \cite{GrKazSat2022} that a discrete Shilnikov attractor exists in its phase space \cite{GonGonShil2012, GonGonKazTur2014, GonGon2016}.
Later, other ways were found to provide chaotic dynamics in such systems, either by an inclusion of higher harmonics \cite{BickTimPaulRathAsh2011, CluPo2020} in the coupling function or by using previously mentioned non-pairwise interactions between phases \cite{BAR16}.

The case of $N = 4$ oscillators is the most remarkable -- this is the minimal number of oscillators in the system of globally connected identical phase oscillators \eqref{eq:KurTypeGlobalIdentical} that allows any non-trivial (including chaotic) dynamics.
This follows from the fact that the system describing the dynamics of $N$ oscillators could be reduced to $N-1$-dimensional system of ODEs: if $N-1 = 2$, then the reduced system has trivial planar dynamics due to Poincar\'{e}-Bendixson theory. 
In \cite{AOWT07} it was noted that although the introduction of the second harmonic can lead to an emergence of chaos for $N \geqslant 5$ oscillators, there is no numerical evidence for its presence when $N = 4$.
In subsequent works, introducing higher harmonics \cite{BickTimPaulRathAsh2011} into the coupling function or non-pairwise interactions \cite{BAR16} lead to new examples of chaotic dynamics in systems of four oscillators.
A vast literature review in recent papers \cite{burylko2020collective, AshBickRodr2022} still does not mention that there are any numerical evidences of chaotic dynamics in a system of four identical globally coupled phase oscillators with a biharmonic coupling function.
This work is based on the observations presented in \cite{BAR16, GO18}.
In these papers it was noted that for the system with non-pairwise interactions its chaotic attractors seem to be organized by heteroclinic cycles that include only equilibria belonging to invariant planes of this system.
These heteroclinic cycles include both trajectories belonging to invariant planes and trajectories outside them.
In this paper we develop a numerical procedure for searching for such heteroclinic cycles and apply it to find examples of chaotic dynamics in this system.
The paper consists of the following sections.
In Section \ref{sec:SysAndSym} we describe the standard method of reduction to a three-dimensional system of differential equations and list properties related to its symmetries.
Section \ref{sec:HetCyclesAndResults} discusses heteroclinic cycles of Tresser type, whose presence may indicate the existence of chaotic dynamics in the phase space. 
We also show the results of studying the system using the proposed algorithm for searching for such heteroclinic cycles.
We give examples of chaotic attractors that we found near the points in the parameter space at which these cycles exist.
Section \ref{sec:Chaos} demonstrates the scenarios of the transition to chaos that we have observed.
The \hyperref[sec:appendix]{Appendix} describes the algorithm for the numerical search of heteroclinic cycles of the Tresser type.

\section{The description of a system and its symmetries}
\label{sec:SysAndSym}
The starting point of our study is the following system \cite{AOWT07} of four globally coupled identical phase oscillators:
\begin{equation}
\label{eq:SystA4d}
\dot{\phi}_n = \omega + \frac{1}{4} \sum_{m=1}^4 g(\phi_n - \phi_m), \;\; n = \overline{1, 4},
\end{equation}
where $g(\varphi) = -\sin( \varphi +\alpha)+r\sin(2 \varphi +\beta)$ is the biharmonic coupling function. 
These equations have remarkable additional properties \cite{AshSwift1992, ABB16, burylko2020collective} due to their special structure (symmetric use of variables, dependence on phase differences).
The system \eqref{eq:SystA4d} has the following symmetries 
\begin{itemize}
\item {if $(\phi_1(t),\phi_2(t),\phi_3(t),\phi_4(t))$ is a solution, then $(\phi_{\sigma(1)}(t),\phi_{\sigma(2)}(t),\phi_{\sigma(3)}(t),\phi_{\sigma(4)}(t))$ is also a solution for any permutation $\sigma \in S_4$, where $S_4$ is a permutation group of 4-element set;}
\item {if $(\phi_1(t),\phi_2(t),\phi_3(t),\phi_4(t))$ is a solution, then $(\phi_1(t) + \Phi,\phi_2(t) +\Phi,\phi_3(t) +\Phi,\phi_4(t) +\Phi)$ is also a solution for any $\Phi \in \mathbb{R}$;}
\item {if $(\phi_1(t),\phi_2(t),\phi_3(t),\phi_4(t))$ is a solution, then $(\phi_1(t) + 2\pi k_1 ,\phi_2(t) + 2\pi k_2,\phi_3(t) + 2\pi k_3,\phi_4(t) + 2\pi k_4)$ is also a solution for any $(k_1, k_2, k_3,k_4) \in \mathbb{Z}^4$. }
\end{itemize}
The first and the third symmetries force hyperplanes $\lbrace \phi_i=\phi_j $ mod $ 2\pi \rbrace$ to be invariant.
Because of that the ordering of phases does not change in time: any permutation $\sigma \in S_4$ and vector $(k_1, k_2, k_3, k_4)\in \mathbb{Z}^4$ correspond to an invariant region defined by the inequalities  
\begin{equation} 
\label{eq:domains}
2\pi k_1 +\phi_{\sigma(1)}(t)\leqslant 2\pi k_2 +\phi_{\sigma(2)}(t)\leqslant 2\pi k_3 + \phi_{\sigma(3)}(t) \leqslant 2\pi k_4 +\phi_{\sigma(4)}(t) \leqslant 2\pi (k_1+1) +\phi_{\sigma(1)}(t).
\end{equation}
All such regions can be obtained from each other by applying some transformation from the symmetry group, hence dynamics in each of these regions is absolutely the same. 
Thus, we can pick any of these regions as a representative for the others.

A standard approach for studying such systems is passing to a reduced system of equations for phase differences $\psi_n = \phi_n - \phi_1.$
The reduced system is described by the following equations 
\begin{equation}
\label{eq:RedusedA4d}
\begin{gathered}
\dot{\psi}_1 = 0,\\
\dot{\psi}_n = \Psi_n (\psi_2, \psi_3, \psi_4) =  \frac{1}{4}\sum\limits_{m=1}^4 \left \lbrack g(\psi_n - \psi_m) -g(-\psi_m) \right \rbrack, n= 2,3,4.
\end{gathered}
\end{equation}
From $\dot{\psi}_1(t) \equiv 0$ and $\psi_1 (0) = 0$ follows that $\psi_1 (t) \equiv 0$. Thus, the equation for $\psi_1$ can be discarded and variables $\psi_2, \psi_3, \dots, \psi_N$ alone describe the behaviour of system \eqref{eq:RedusedA4d}.

\begin{figure}[!thb]
\centering
\includegraphics[width=0.4\textwidth]{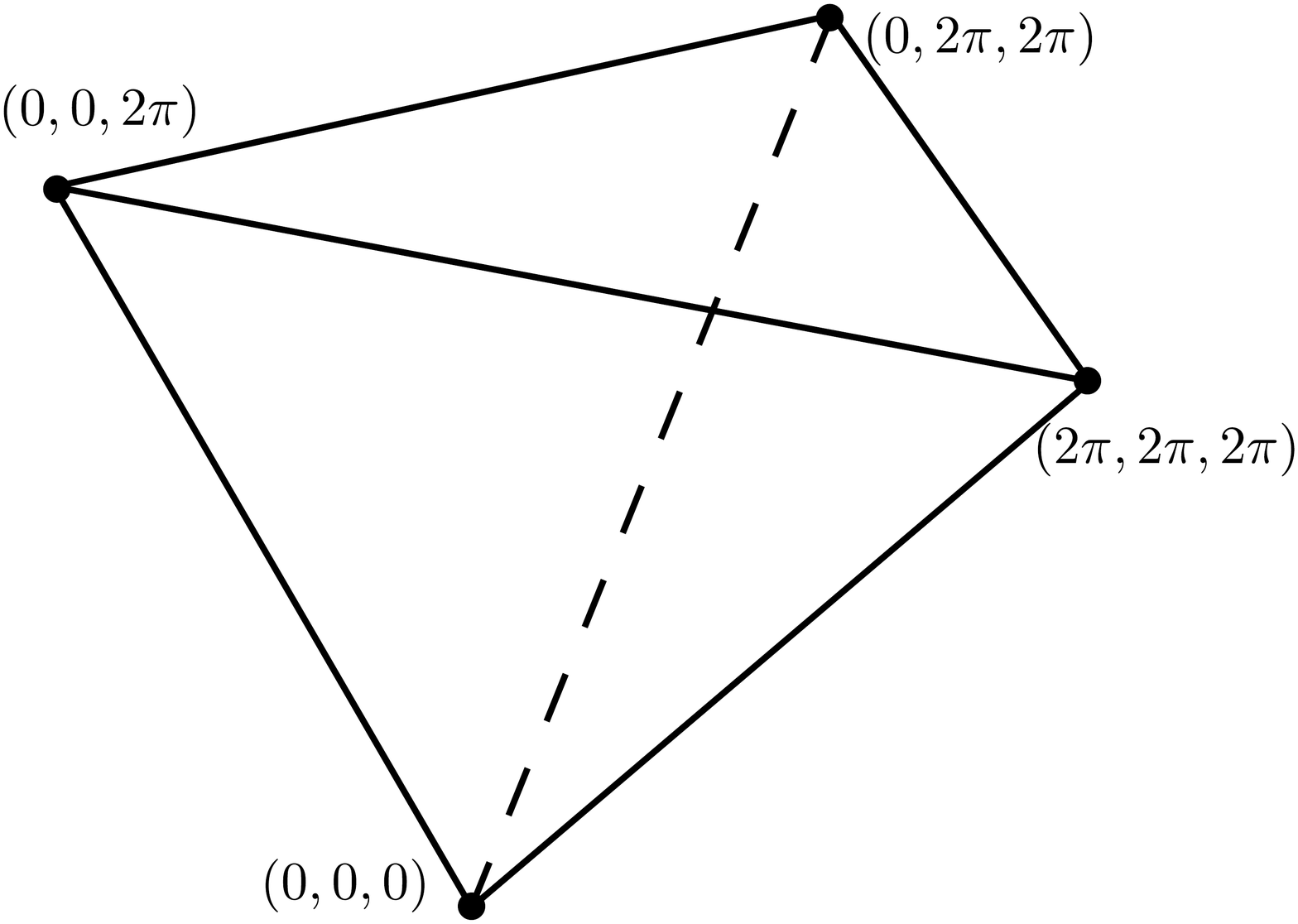}
\caption{\label{fig:CIR} The canonical invariant region $\mathcal{C}$ of the system \eqref{eq:RedusedA4d}}
\end{figure}

Let us note an important geometric feature of the reduced system, namely, how the invariant regions described by the inequalities \eqref{eq:domains} are transformed by the reduction. 
Passing to phase differences transforms the set $\phi_1\leqslant \phi_2\leqslant \phi_3 \leqslant \phi_4 \leqslant \phi_1+2\pi$ into the set $ 0 \leqslant \psi_2\leqslant \psi_3 \leqslant \psi_4 \leqslant 2\pi$. Following \cite{AshSwift1992, BAR16}, we will call this set a \textit{canonical invariant region} $\mathcal{C}$.
The set $\mathcal{C}$ geometrically is just a tetrahedron in $\mathbb{R}^3$ with its interior (see Fig. \ref{fig:CIR}), so we can define vertices, edges and faces of a canonical invariant region $\mathcal{C}$ in the same fashion as for tetrahedra.
Note that a reduced system \eqref{eq:RedusedA4d} inherits discrete symmetries of system \eqref{eq:SystA4d}.
If we keep convention that $\psi_1=0$, then a permutation $\sigma \in S_4$ and vector $(k_1,k_2,k_3) \in \mathbb{Z}^3$ induce a map  
$$(\psi_2, \psi_3, \psi_4)\mapsto (\psi_{\sigma(2)}-\psi_{\sigma(1)} +2\pi k_1,\psi_{\sigma(3)}-\psi_{\sigma(1)} +2\pi k_2,\psi_{\sigma(4)}-\psi_{\sigma(1)} +2\pi k_3), $$
which is a symmetry of reduced system \eqref{eq:RedusedA4d}.
A canonical invariant region $\mathcal{C}$ has its own symmetry subgroup then, which is generated by a mapping \cite{BAR16}
$$ T: (\psi_2, \psi_3, \psi_4)\mapsto (\psi_3-\psi_2,\psi_4-\psi_2,2\pi - \psi_2). $$
A simple calculation shows that $T^4 = {\rm id}$ and $T^k \neq {\rm id}$ for $k = 1, 2, 3$, where ${\rm id}: (\psi_2, \psi_3, \psi_4) \mapsto (\psi_2, \psi_3, \psi_4)$ is the identity mapping. 
Thus, the symmetry group of the canonical invariant region $\mathcal{C}$ is generated by the mapping $T$ and it is isomorphic to  $\mathbb{Z}_4 = \mathbb{Z} / 4\mathbb{Z}$.

\section{Heteroclinic cycles of system under the study}
\label{sec:HetCyclesAndResults}
\subsection{Key features of heteroclinic cycles}
\label{sec:HetCycles}
\begin{figure}[!thb]
\centering
\includegraphics[width=0.65\textwidth]{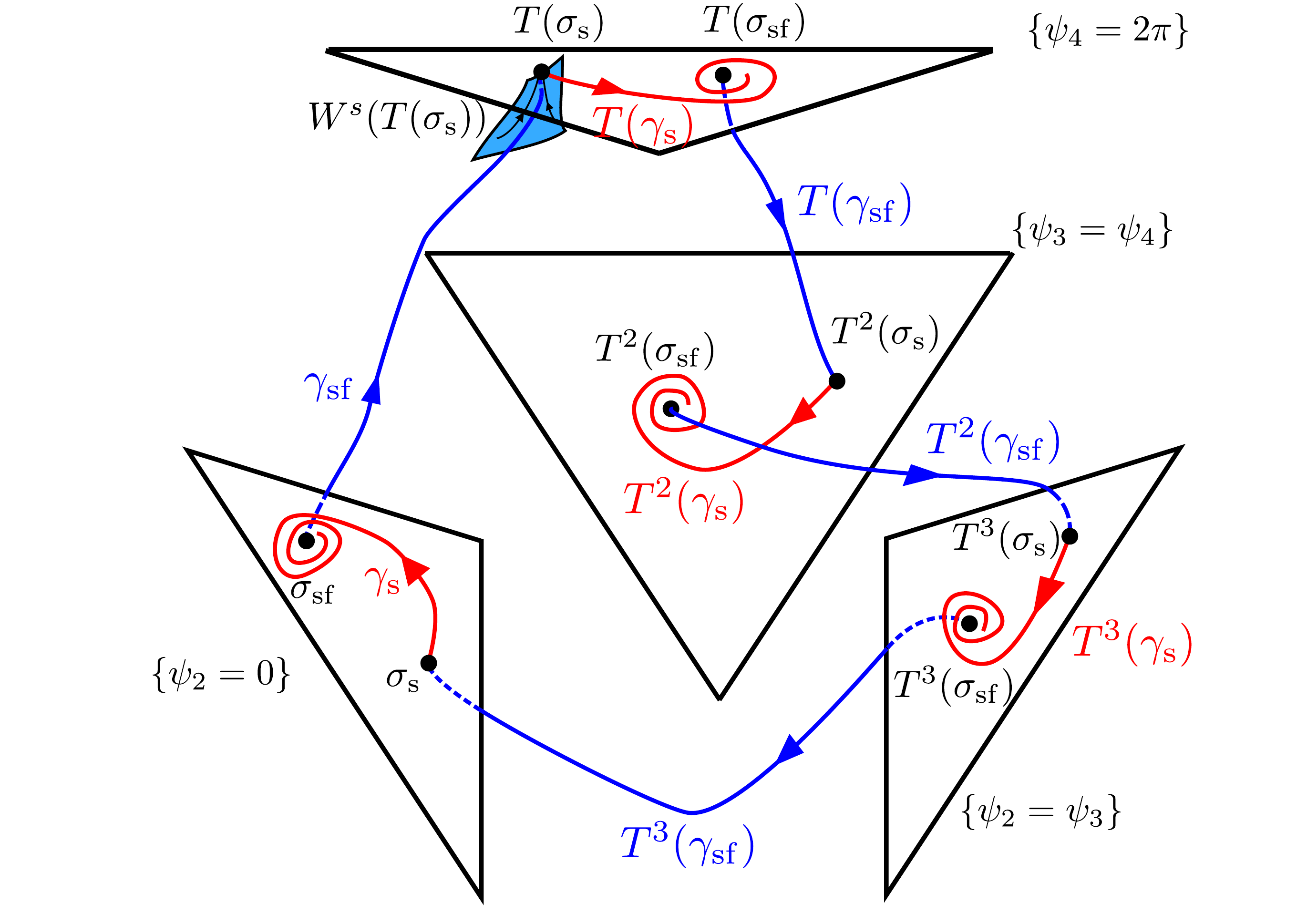}
\caption{\label{fig:HeterClin}Schematic structure of the heteroclinic cycle corresponding to the connecting trajectories $\sigma_{\rm s} \rightarrow \sigma_{\rm sf} \rightarrow T(\sigma_{\rm s})$}
\end{figure}

Following work \cite{GO18}, we focus on finding and studying heteroclinic cycles similar to those depicted in Fig. \ref{fig:HeterClin}.
Such heteroclinic cycles include equilibrium states $\sigma_{\rm s}$ (saddle) and $\sigma_{\rm sf}$ (saddle-focus) belonging to the same invariant face of the tetrahedron, and some symmetric copy $T^{k}(\sigma_{\rm s})$ that is also located on some invariant face (possibly the same one); the integer $k$ can take any value from $0$ to $3$.
Denote by $\sigma_{\rm s} \rightarrow \sigma_{\rm sf} \rightarrow T^k(\sigma_{\rm s})$ any such heteroclinic cycle, where $\sigma_{\rm s}$ is a saddle equilibrium state with a two-dimensional stable manifold transversal to an invariant face, and $\sigma_{\rm sf}$ is a saddle-focus with a one-dimensional unstable invariant manifold.
The arrows in this notation mean that there is a heteroclinic trajectory between adjacent equilibrium states.
Note that to find any heteroclinic cycle of this type, it is sufficient to know only two heteroclinic trajectories: from $\sigma_{\rm s}$ to $\sigma_{\rm sf}$ and from $\sigma_{\rm sf}$ to $T^k(\sigma_{\rm s})$.
Indeed, if the trajectory $\gamma_{\rm s}$ is heteroclinic with respect to the equilibria $\sigma_{\rm s}$ and $\sigma_{\rm sf}$, then the trajectory $T^m(\gamma_{\rm s})$ is heteroclinic between equilibria $T^m(\sigma_{\rm s})$ and $T^m(\sigma_{\rm sf})$ for any $m$: this is a direct consequence of the fact that the map $T^m$ is continuous and it is a symmetry of the system.
A similar conclusion can be drawn for the heteroclinic trajectory $\gamma_{\rm sf}$ connecting the equilibrium states $\sigma_{\rm sf}$ and $T^k(\sigma_{\rm s})$: the trajectory $T^m(\gamma_{\rm sf})$ is heteroclinic between equilibrium states $T^m(\sigma_{\rm sf})$ and $T^{m+k}(\sigma_{\rm s})$.
Then, applying the mapping $T^k$ and its powers to the connecting trajectories $\sigma_{\rm s} \rightarrow \sigma_{\rm sf} \rightarrow T^k(\sigma_{\rm s})$, we obtain the following chain of equilibrium states and connections between them
$$
\begin{array}{l} \sigma_{\rm s} \rightarrow \sigma_{\rm sf} \rightarrow T^k(\sigma_{\rm s}) \rightarrow T^{k}(\sigma_{\rm sf}) \rightarrow T^{2k}(\sigma_{\rm s}) \rightarrow  T^{2k}(\sigma_{\rm sf}) \rightarrow T^{3k}(\sigma_{\rm s}) \rightarrow T^{3k}(\sigma_{\rm sf}) \rightarrow T^{4k}(\sigma_{\rm s}) = \sigma_{\rm s},
\end{array}
$$
which always loops due to the fact that $T^{4k} = (T^4)^k$ and $T^4$ is the identical transformation.
Note that the resulting heteroclinic cycle can also consist of a smaller number of equilibrium states: two, if $k = 0$, and four, if $k = 2$.
If $k = 1$ or $k = 3$, then the heteroclinic cycle consists of eight equilibria, and the value of $k$ affects only the order in which the equilibria appear on the faces of the invariant tetrahedron.
Fig.~\ref{fig:HeterClin} shows a diagram of such heteroclinic cycle for the case $\sigma_{\rm s} \rightarrow \sigma_{\rm sf} \rightarrow T(\sigma_{\rm s})$.
It is known from the work of Tresser \cite{Tresser1984} that under certain conditions the presence of heteroclinic cycle similar to the described above entails the existence of a countable number of closed trajectories in its vicinity, and, hence, complex non-trivial dynamics.
One of these conditions is the condition on the product of saddle values $p > 1$, which is calculated by the formula
$$p\; \; \; \; = \; \; \prod\limits_{O_j \text{\, -- saddle-focus}} -\frac{\lambda^{\rm sf}_j}{\rho^{\rm sf}_j} \cdot \prod\limits_{O_j \text{\, -- saddle}} -\frac{\lambda^{\rm s}_{1, j}}{\lambda^{\rm s}_{2,j}},$$
where $O_1, \; O_2, \; \dots \; ,\; O_n$ is a some sequence of saddles and saddle-foci of the three-dimensional system, organized into a heteroclinic cycle.
In this formula we assume that if the equilibrium state $O_j$ is a saddle-focus, then its eigenvalues are $$\lambda^{\rm sf}_j > 0; \rho^{\rm sf}_j \pm i\omega^{\rm sf}_j, \rho^{\rm sf}_j < 0. $$
If the equilibrium state $O_j$ is a saddle, then its eigenvalues satisfy the inequality $$\lambda^{\rm s}_{3, j} < \lambda^{\rm s}_{2, j} < 0 < \lambda^{\rm s}_{1, j} .$$
Note that this formula is simplified for the heteroclinic cycles we are considering.
Without loss of generality, let us analyze this using the example of the already mentioned $\sigma_{\rm s} \rightarrow \sigma_{\rm sf} \rightarrow T(\sigma_{\rm s})$ heteroclinic cycle .
In this case, the saddle-foci included in the heteroclinic cycle are the equilibrium states $\sigma_{\rm sf}$, $T(\sigma_{\rm sf})$, $T^2(\sigma_{\rm sf})$, and $T^3(\sigma_{\rm sf})$; the saddle equilibria are $\sigma_{\rm s}$, $T(\sigma_{\rm s})$, $T^2(\sigma_{\rm s})$, and $T^3(\sigma_{\rm s})$.
Since $T$ is also a smooth mapping, the equilibria $\sigma_{\rm sf}$ and $T^n(\sigma_{\rm sf})$ have the same set of eigenvalues for any $n$; the same holds for $\sigma_{\rm s}$ and $T^n(\sigma_{\rm s})$.
We will assume that for eigenvalues of the saddle $\sigma_{\rm s}$ holds $$\lambda^{\rm s}_1 > 0; \lambda^{\rm s}_3 <\lambda^{\rm s}_2 < 0. $$  
Also, we will assume that eigenvalues of the saddle-focus $\sigma_{\rm sf}$ satisfy $$\lambda^{\rm sf} > 0; \rho^{\rm sf} \pm i\omega^{\rm sf}, \rho^{\rm sf} < 0. $$
Using these assumptions we get that
$${p} = \left ( -\frac{\lambda^{\rm sf}}{\rho^{\rm sf}} \right )^4 \cdot \left (-\frac{\lambda^{\rm s}_1}{\lambda^{\rm s}_2} \right )^{4} = \tilde{p}^{\,4},$$
where
\begin{equation} 
\label{eq:ptild}
\tilde{p} = \frac{\lambda^{\rm sf}}{\rho^{\rm sf}} \cdot \frac{\lambda^{\rm s}_1}{\lambda^{\rm s}_2}. 
\end{equation}
Obviously, $p > 1$ if and only if $\tilde{p} > 1$.
Using the introduced notation and the previous observations, we write down the necessary conditions from the formulation of the Tresser theorem that are most important for us:
\begin{enumerate}
\item There are parameter values for which there is a configuration of saddles and saddle-foci on invariant faces that allows a heteroclinic cycle. As was already described earlier, the saddle $\sigma_{\rm s}$ must have a two-dimensional stable manifold transversal to the invariant face; it also must be a saddle for the restriction of the system to the invariant face. The saddle-focus $\sigma_{\rm sf}$ is required to be a stable focus for the restriction of the system to an invariant face, and, consequently, its one-dimensional unstable manifold has to be transversal to it.
\item For the same values of the parameters, the presence of heteroclinic trajectories $\gamma_{\rm s}$ and $\gamma_{\rm sf}$ is required. This guarantees the existence of a heteroclinic cycle that includes the equilibrium states $\sigma_{\rm s}$ and $\sigma_{\rm sf}$, as well as some of their symmetrical copies.
\item The eigenvalues of these equilbria satisfy  conditions described in Tresser's theorem \cite{Tresser1984}.
Satisfying all the above conditions entails the existence of a Smale horseshoe in the neighborhood of the heteroclinic cycle and, therefore, nontrivial complex dynamics. In our case, this is equivalent to the condition $\tilde{p} > 1$.
\end{enumerate}
Note that the necessary conditions for applying Tresser's theorem also include requirements related to the intersection geometry of invariant manifolds.
In our proposed approach, we check only the three conditions mentioned above. Despite checking the weaker set of conditions, this approach can still be used to narrow the search region in the parameter space and identify sub-regions where chaotic dynamics is possible.
These subdomains can be further investigated, for example, by methods based on the calculation of Lyapunov exponents, which is demonstrated in Section \ref{sec:ResultsApprox}.
The verification of these conditions forms the basis of the algorithm for searching for approximate heteroclinic cycles, the key features of which are described in the \hyperref[sec:appendix]{Appendix}.

\subsection{Results of numerical search for heteroclinic cycles}
\label{sec:ResultsApprox} 
\begin{figure}
\begin{center}
\includegraphics[height = 8cm]{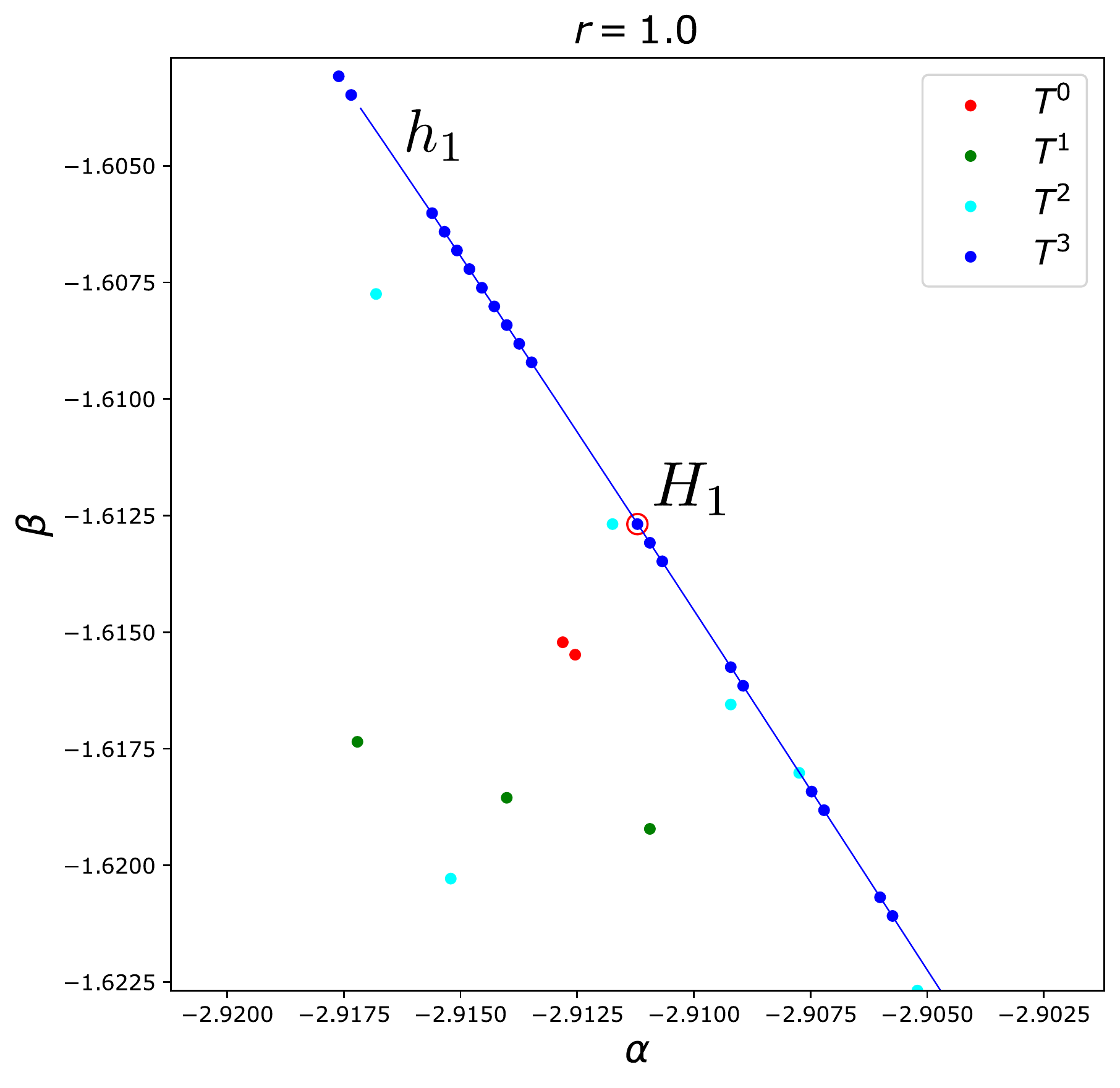}
\caption{\label{fig:HetMap} Map of heteroclinic cycles. The dots mark the values of the parameters at which a heteroclinic cycle was found in the system using the proposed approach. The color indicates the connections $\sigma_{\rm s} \rightarrow \sigma_{\rm sf} \rightarrow T^k(\sigma_{\rm s})$ that generate the cycle. 
}
\end{center}
\end{figure}

\begin{figure}
\begin{center}
\includegraphics[height = 8cm]{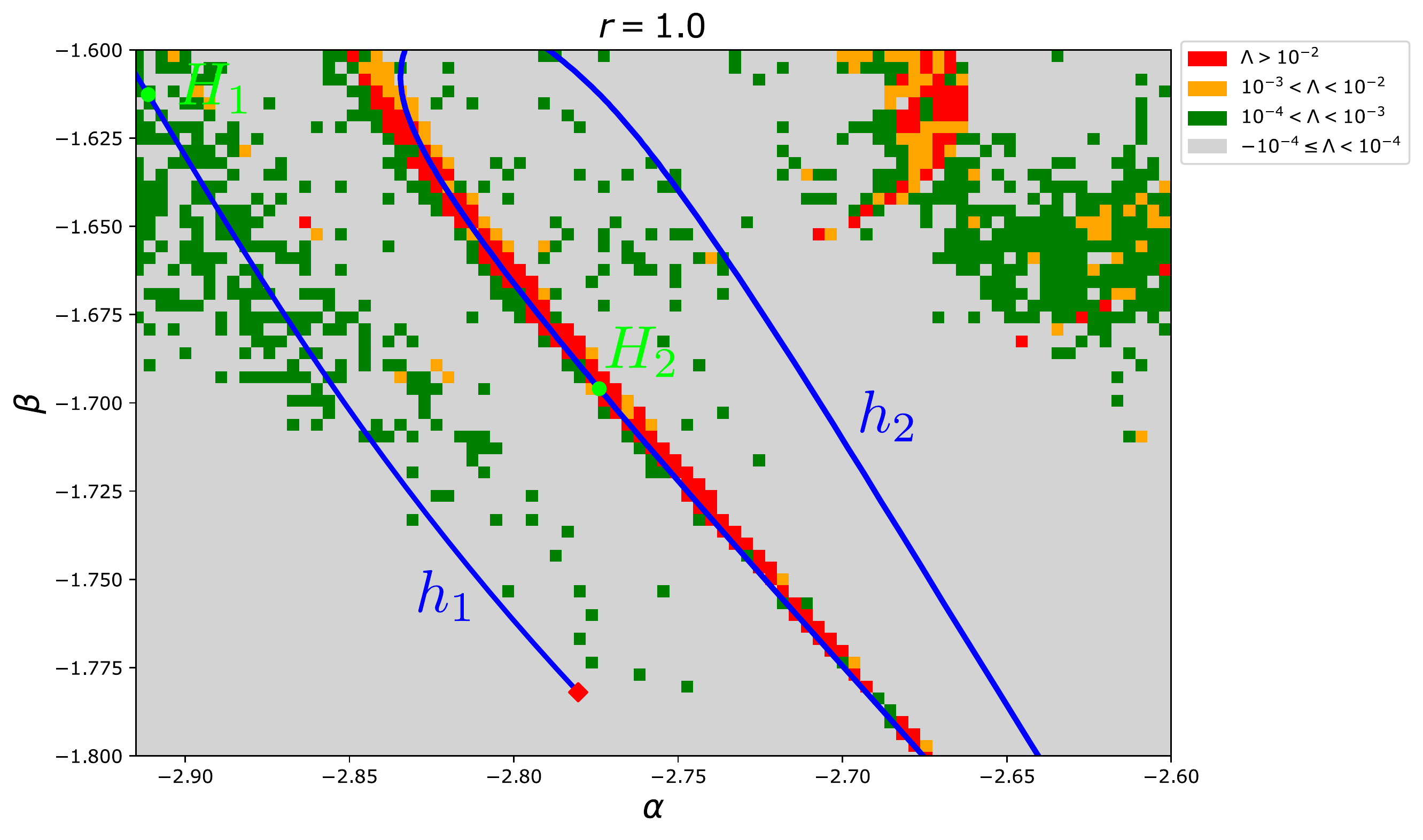}
\caption{\label{fig:LyapMap} A map of the largest Lyapunov exponent superimposed with curves of the existence of heteroclinic cycles. Bright green circles mark the points $H_1$ and $H_2$ where heteroclinic cycles were found by an algorithm. These points were used to compute the blue curves of heteroclinic cycles ($h_1$ and $h_2$) using MATCONT. The red rhombus is the point at which the saddle-focus of the heteroclinic cycle becomes a saddle with purely real eigenvalues.}
\end{center}
\end{figure}

Fig.~\ref{fig:HetMap} shows a map of approximate heteroclinic cycles obtained using the proposed algorithm.
For a fixed value of the parameter $r = 1$, a uniform grid $151 \times 151$ is constructed on the plane of the parameters $\alpha$ and $\beta$. At each grid node we run an algorithm, and
the dots mark the values of the parameters for which a heteroclinic cycle was found.
The separatrix between the saddle-focus and the saddle found at the point $H_1$ was used as an initial approximation in the MATCONT package \cite{DhGoKuMeiSau}, which implements the procedure for the numerical continuation of a heteroclinic trajectory with respect to parameters.
This gives a curve $h_1$ in Fig.~\ref{fig:HetMap} that agrees with other blue points found by our algorithm.
We specifically check that all other conditions for the existence of the heteroclinic cycle are satisfied along this curve.
Thus, curve $h_1$ represents not only the existence of the heteroclinic trajectory between saddle-focus and saddle equilibria, but the existence of the whole heteroclinic cycle satisfying the requirements that we have mentioned in Section \ref{sec:HetCyclesAndResults}.
Fig.~\ref{fig:LyapMap} also shows this curve $h_1$ superimposed with the map of the largest Lyapunov exponent, which is computed using DynamicalSystems.jl package \cite{Datseris2018}.
The red rhombus is the point at which the saddle-focus of the heteroclinic cycle becomes a saddle with purely real eigenvalues.
The curve $h_1$ is shown only up to this point since fulfilling the conditions of the Tresser theorem requires a presence of at least one saddle-focus in a heteroclinic cycle.
It can be seen that, next to the heteroclinic curve $h_1$, there are parameter values corresponding to weak chaos (the largest Lyapunov exponent $\Lambda \approx 10^{-4}\text{-}10^{-3}$).
The algorithm also has found the point $H_2$ in the sub-region of the parameter plane containing points with stronger chaos (marked in red), where $\Lambda \gtrsim 10^{-3}\text{-}10^{-2}$.
The saddle-focus separatrix found at this point was also used for computing the curve of the heteroclinic cycle's existence $h_2$.


\begin{figure}[!thb]
\centering
{
\begin{minipage}[t]{0.28\linewidth}
\center{\includegraphics[width=1\linewidth]{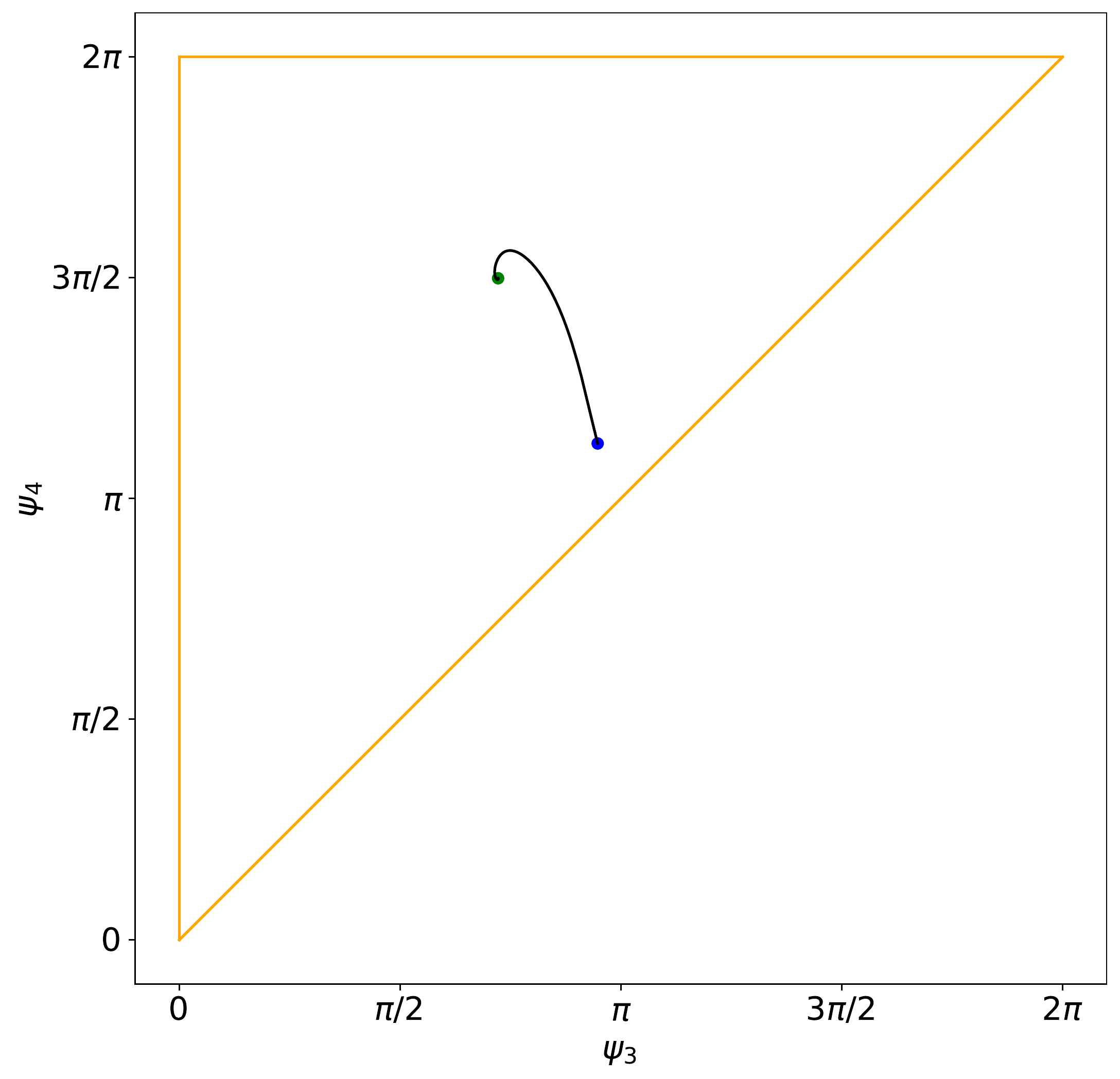}} \\a) 
\end{minipage}
}
{
\begin{minipage}[t]{0.35\linewidth}
\center{\includegraphics[width=0.9\linewidth]{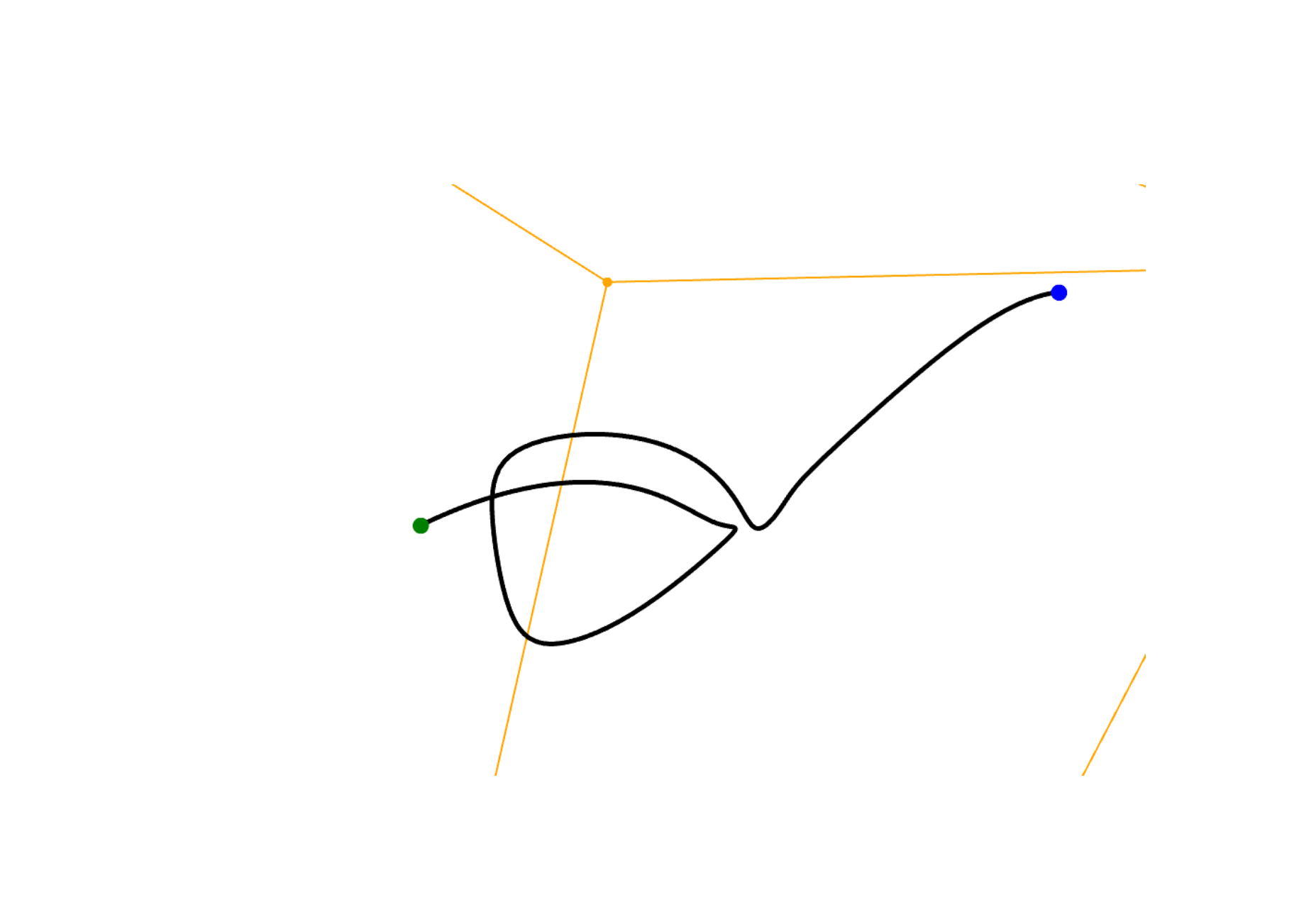}} \\b) 
\end{minipage}
}
{
\begin{minipage}[t]{0.25\linewidth}
\center{\includegraphics[width=0.9\linewidth]{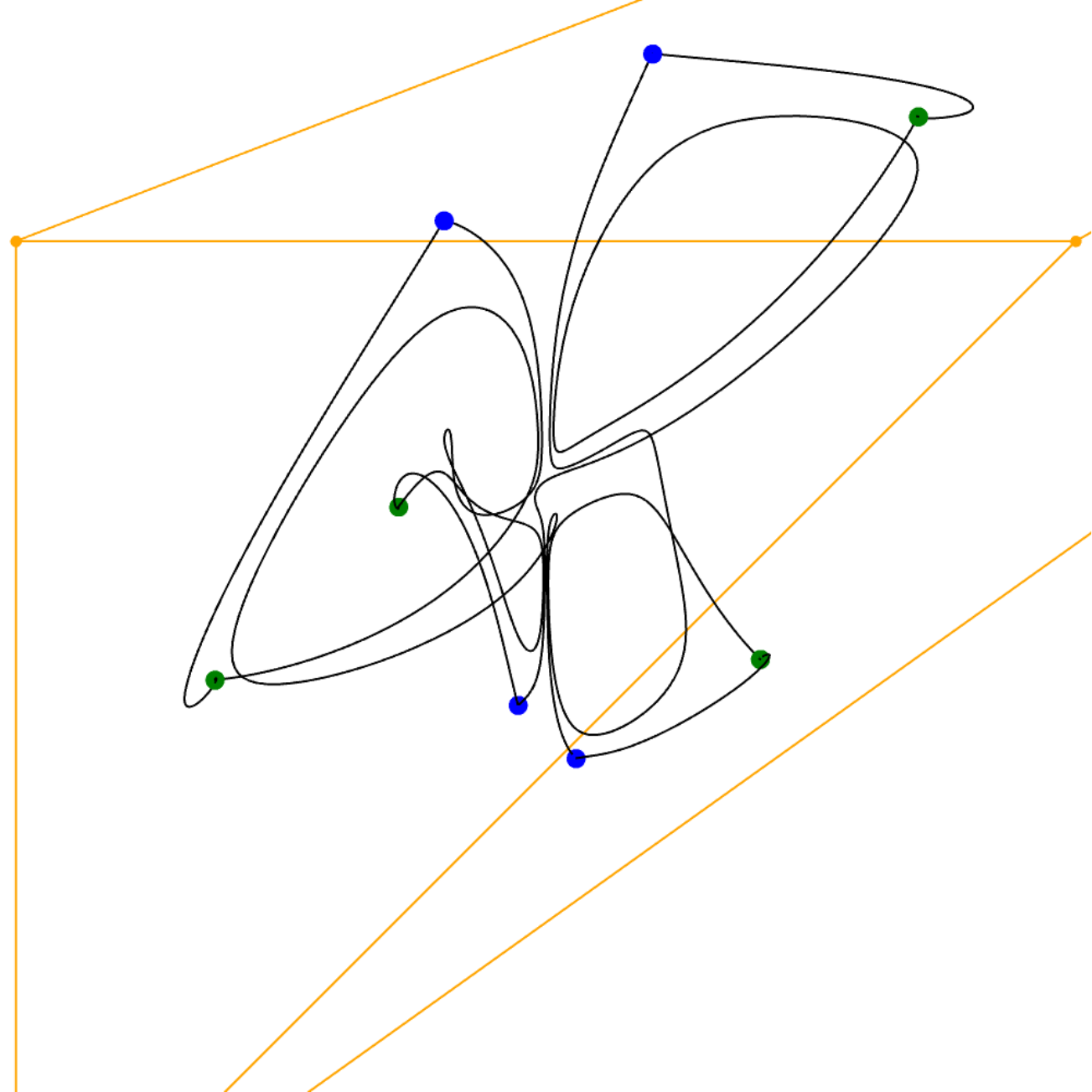}} \\c)
\end{minipage} 
}
\\
\centering
\begin{minipage}[t]{0.49\linewidth}
\center{\includegraphics[height=4cm]{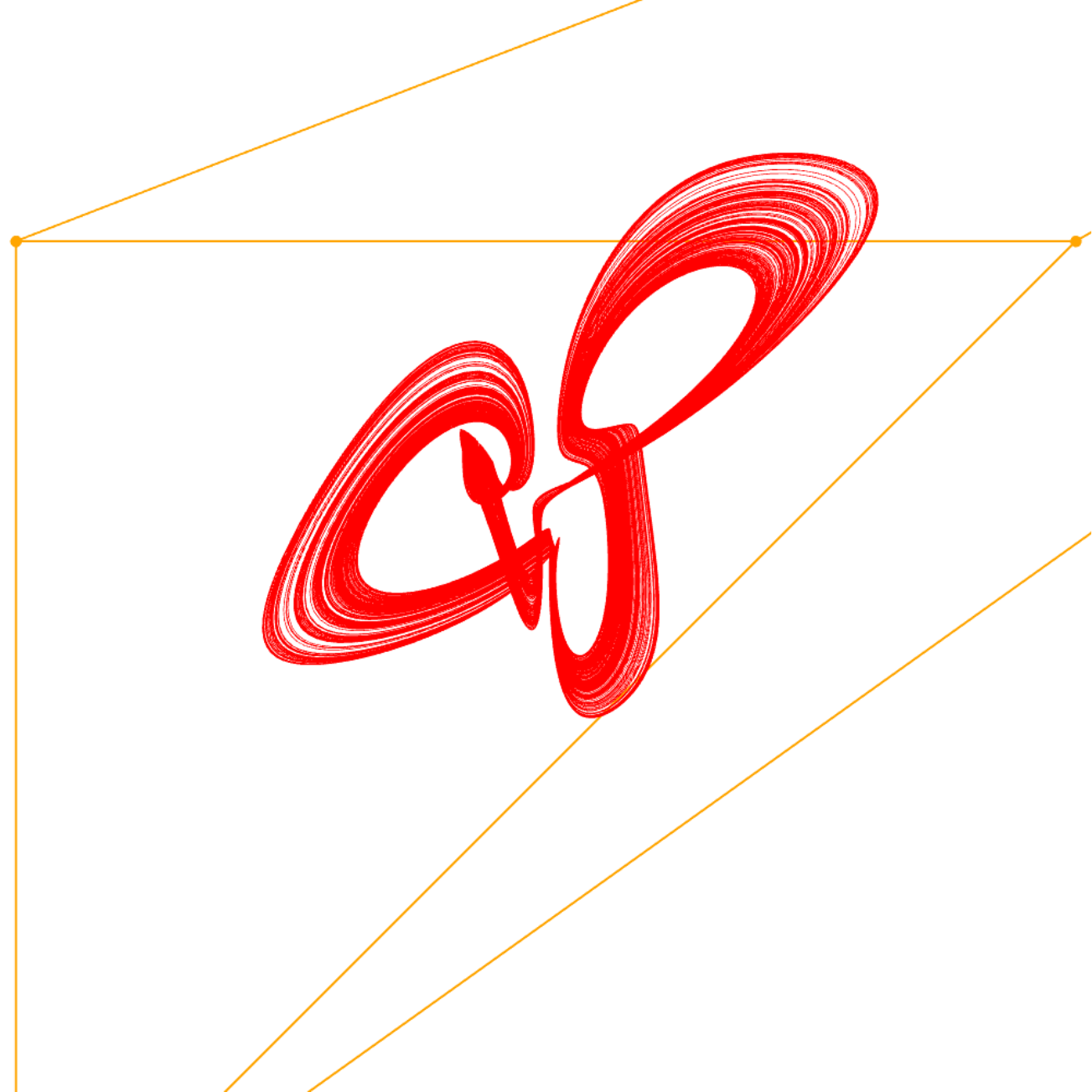}} \\d) 
\end{minipage}
\begin{minipage}[t]{0.50\linewidth}
\center{\includegraphics[height=4cm]{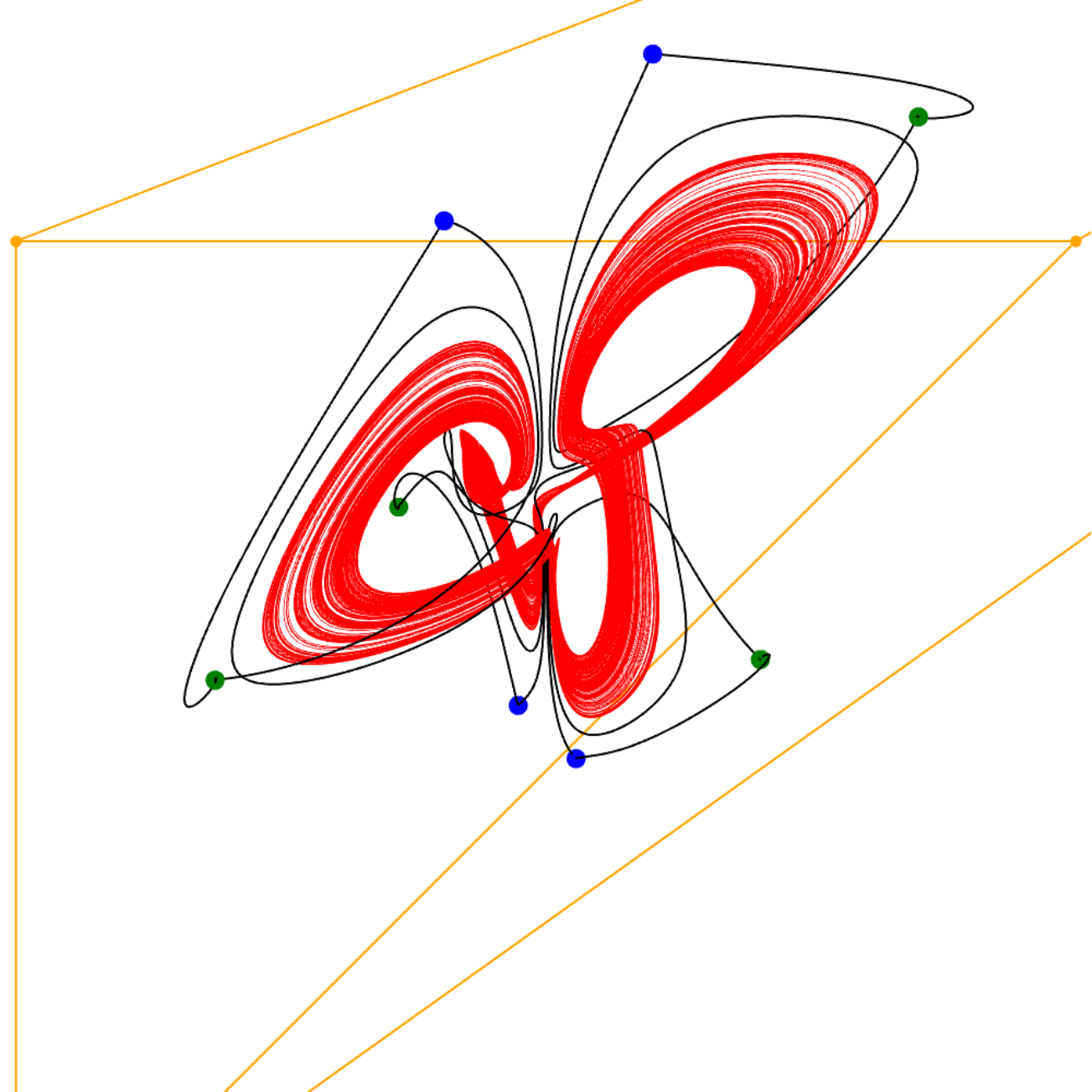}} \\e) 
\end{minipage}
\caption{\label{fig:chaosAttrAndHeter}An example of a numerically found heteroclinic cycle and a chaotic attractor near it for parameters $r = 1,\, \alpha = -2.774,\, \beta = -1.696$ (the point $H_2$). The saddle-focus $\widetilde \sigma_{\rm sf}$ is marked in green, the saddle $\widetilde \sigma_{\rm s}$  is marked in blue. a) Separatrix $\widetilde \gamma_{\rm s}$; b) separatrix $\widetilde \gamma_{\rm sf}$; c) heteroclinic cycle, green dots -- saddle foci, blue dots -- saddles; d) chaotic attractor; e) the relative position of the chaotic attractor (in red) and the heteroclinic cycle (in black).}
\end{figure}

Let us also demonstrate an example of a heteroclinic cycle found by the algorithm.
At the parameter values $r = 1, \; \alpha = -2.774, \; \beta = -1.696$ the algorithm has found equilibria $\widetilde \sigma_{\rm s}$ and $\widetilde \sigma_{\rm sf}$ on the face $ \psi_2 = 0$ of the invariant tetrahedron $\mathcal{C}$. 
The saddle $\widetilde \sigma_{\rm s}$ has coordinates 
$$\psi_2 = 0,\, \psi_3 = 2.9762588812635737,\, \psi_4 = 3.5328428198747828, $$
while the saddle-focus $\widetilde \sigma_{\rm sf}$ is a point with coordinates
$$\psi_2 = 0,\, \psi_3 = 2.268022700942448,\, \psi_4 = 4.708575957937916.$$
Substituting the coordinates of the equilibrium states $\widetilde \sigma_{\rm s}$ and $\widetilde \sigma_{\rm sf}$ into the analytically found Jacobi matrix and using numerical methods for finding the eigenvalues of these matrices gives the sets
$$\widetilde \lambda^{\rm s}_1 = 0.3478055487874978, \widetilde \lambda^{\rm s}_2 = -0.4155758664729577, \widetilde \lambda^{\rm s}_3 = -1.4459936877507733,$$
and
$$\widetilde \lambda^{\,\rm sf}_1 = 0.7280650181528661,$$
$$ \widetilde \lambda^{\,\rm sf}_{2,3} = \widetilde{\rho}^{\;\rm sf} \pm i\, \widetilde{\omega}^{\,\rm sf} =  -0.5365352371825014 \pm 0.6665082967975146i$$
respectively.
If we consider the restriction of the system to the invariant plane $\psi_2 = 0$, then the saddle $\widetilde \sigma_{\rm s}$ possesses a pair of eigenvalues $\widetilde \lambda^{\rm s}_1$ and $ \widetilde \lambda^{\rm s}_3$; the saddle-focus $\widetilde \sigma_{\rm sf}$ has eigenvalues  $\widetilde \lambda^{\rm sf}_2$ and $ \widetilde \lambda^{\rm sf}_3$.
Thus, on the invariant plane equilibrium $\widetilde \sigma_{\rm s}$ is a saddle, and $\widetilde \sigma_{\rm sf}$ is a stable focus.
For the equilibrium states $\widetilde \sigma_{\rm s}$ and $\widetilde \sigma_{\rm sf}$ the value of $\tilde{p} = 1.1356855558294372$, which satisfies the condition on the eigenvalues specified in Section \ref{sec:HetCycles}.
The remaining conditions associated with the existence of approximate heteroclinic connections are also satisfied: Fig.~\ref{fig:chaosAttrAndHeter}a shows an unstable separatrix $\widetilde \gamma_{\rm s}$ of saddle $\widetilde \sigma_{\rm s}$ tending to $\widetilde \sigma_{\rm sf}$, which corresponds to a structurally stable saddle-sink connection.
The approximation of the unstable separatrix $\widetilde \gamma_{\rm sf}$ of the saddle-focus $\widetilde \sigma_{\rm sf}$ passes quite close (Fig.~\ref{fig:chaosAttrAndHeter}b) to the saddle $T^3(\widetilde \sigma_{\rm s})$ with coordinates $$\psi_2 = 2.750343716402557, \psi_3 = 2.750343716402557, \psi_4 = 5.726603217855624.$$
Here, the integration of the phase trajectory was stopped at the moment when the point on the phase trajectory was at a distance $\epsilon = 0.009$ from the saddle $T^3(\widetilde \sigma_{\rm s})$.
Fig. \ref{fig:chaosAttrAndHeter}c depicts a heteroclinic cycle formed by $\widetilde \sigma_{\rm s} \rightarrow \widetilde \sigma_{\rm sf} \rightarrow T^3(\widetilde \sigma_{\rm s})$ connections.
At the same parameter values, the system has a chaotic attractor (Fig. \ref{fig:chaosAttrAndHeter}d), for which the value of the largest Lyapunov exponent calculated over $10^5$ time units is approximately equal to 0.018.
\begin{figure}[!thb]
\centering
\begin{minipage}[h]{0.30\linewidth}
\center{\includegraphics[height=4cm]{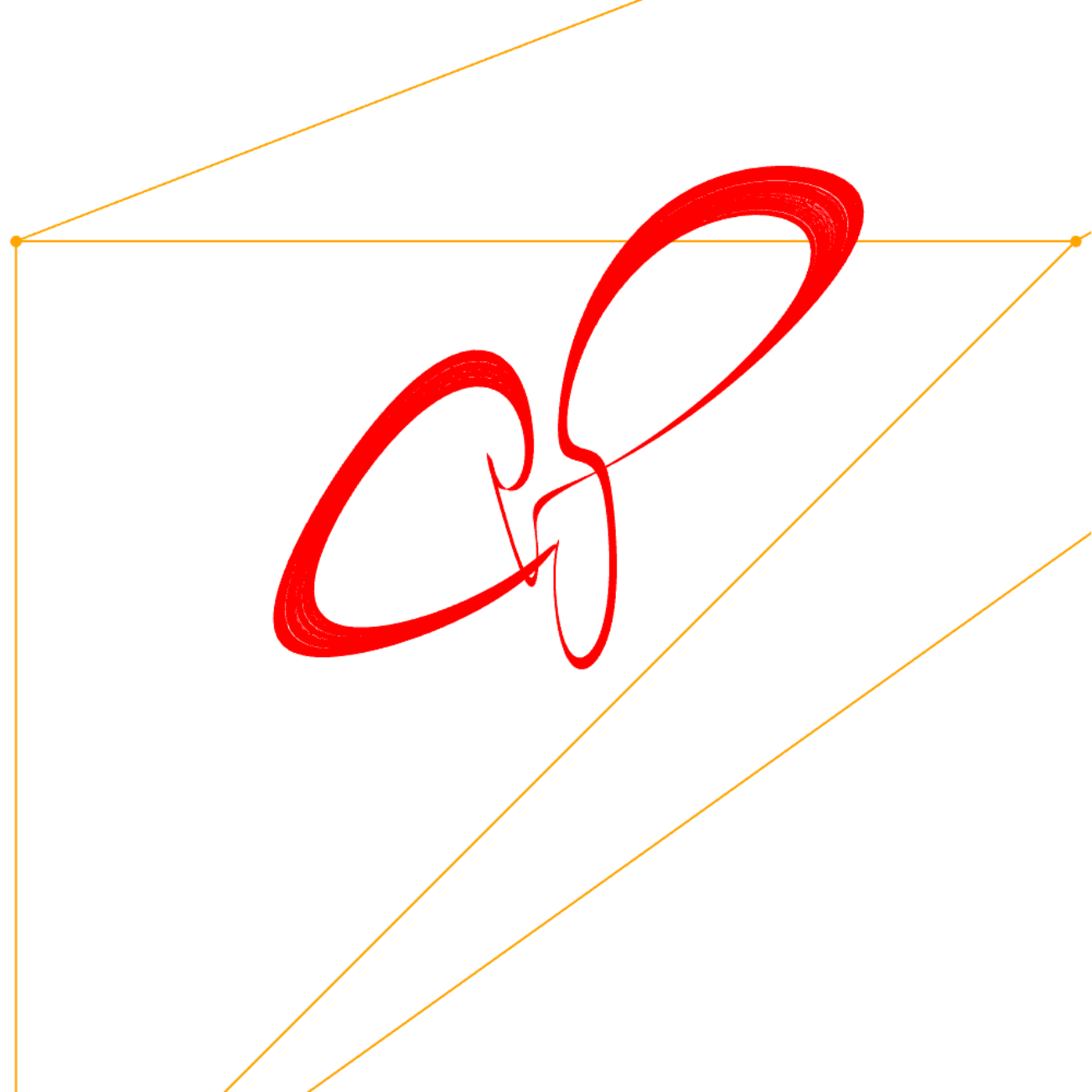}} \\a) 
\end{minipage}
\begin{minipage}[h]{0.30\linewidth}
\center{\includegraphics[height=4cm]{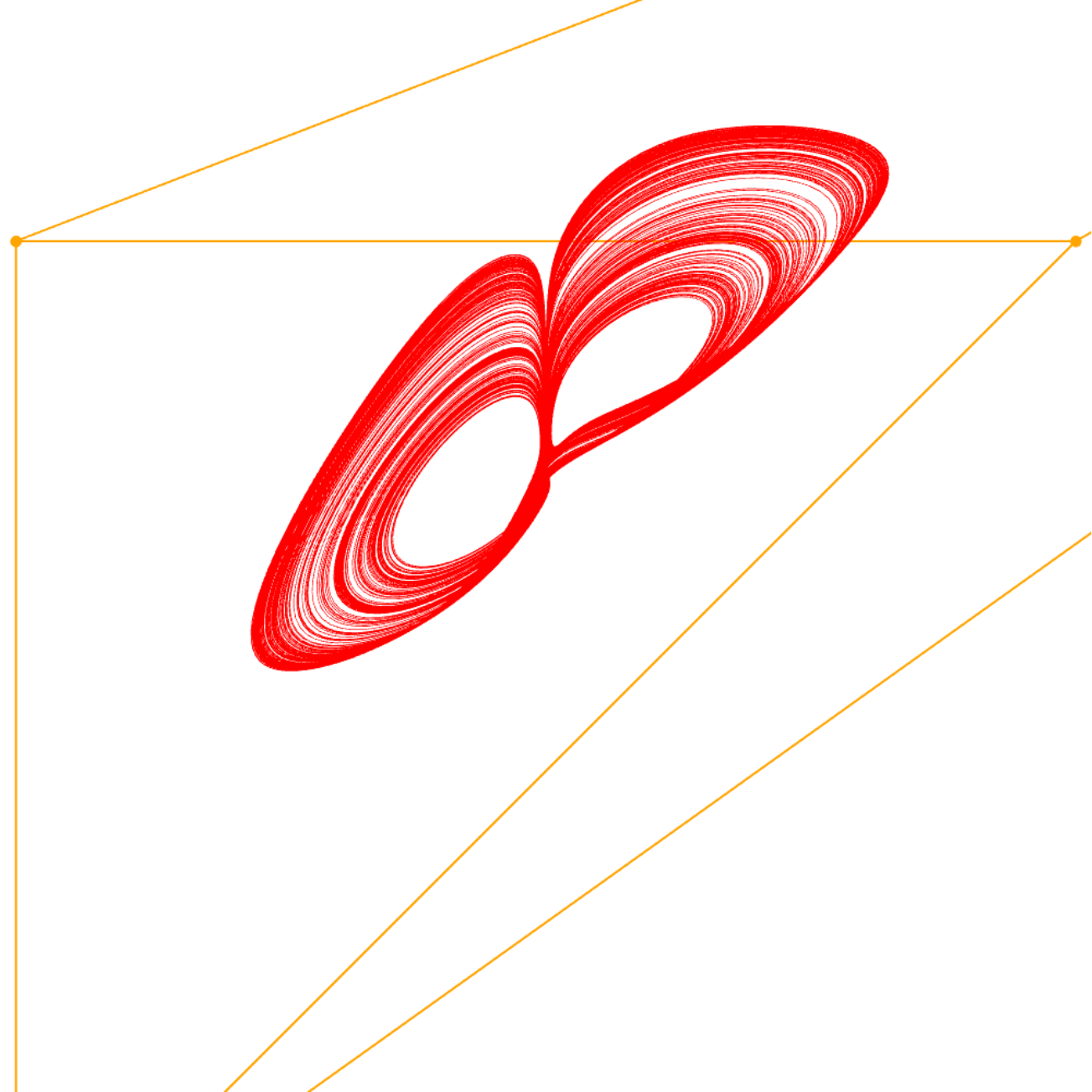}} \\b)
\end{minipage}
\begin{minipage}[h]{0.30\linewidth}
\center{\includegraphics[height=4cm]{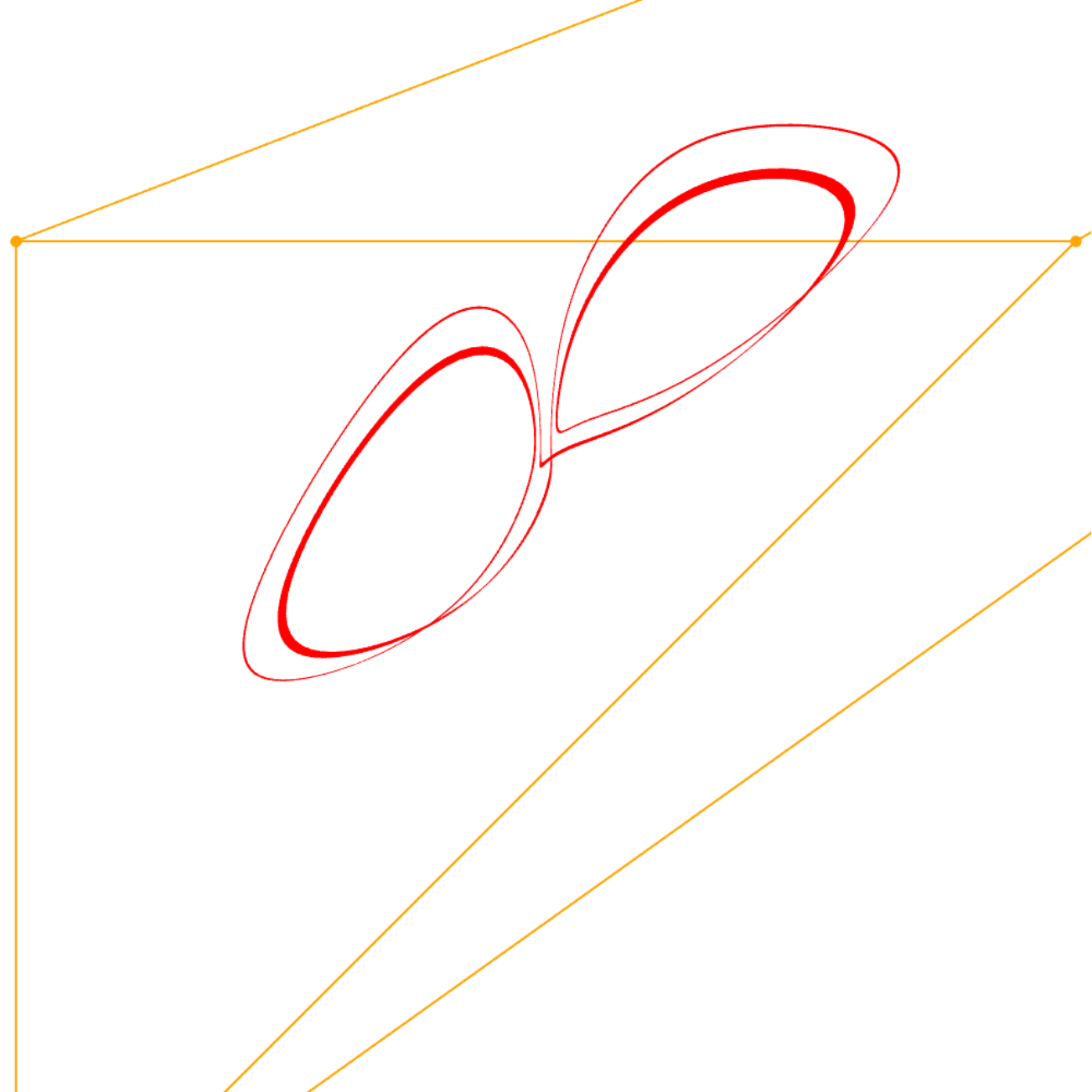}} \\c) 
\end{minipage}
\caption{\label{fig:chaos_example}Other examples of chaotic attractors for $r = 1$ (the corresponding values of the parameters and the value $\Lambda$ of the largest Lyapunov exponent are indicated): a) $\alpha = -2.774,\, \beta = -1.69,\, \Lambda = 0.0117$; b) $\alpha = -2.672,\, \beta = -1.616,\, \Lambda = 0.0135$; c) $\alpha \approx -2.647,\, \beta \approx -1.684,\, \Lambda = 0.0099$.}
\end{figure}

\section{Scenarios of transition to chaos}
\label{sec:Chaos}
Here we describe several scenarios of transition to chaos that we have found.
In both cases we resort to a one-parameter analysis, fixing the values for two out of three parameters, and numerically continue the attractor of interest with respect to the third parameter.
To continue the limit cycles and find their bifurcations, the MATCONT package was used where it was possible.
Otherwise, the attractor was obtained by using a point on the attractor at the previous value of the parameter as the initial condition and discarding the transient process before approaching the attractor (roughly $10^4$ time units).
\subsection{Merging of symmetric limit cycles with subsequent transition to chaos through a cascade of period-doubling bifurcations}
\begin{figure}[!thb]
\begin{minipage}[h]{0.20\linewidth}
\center{\includegraphics[width=1\linewidth]{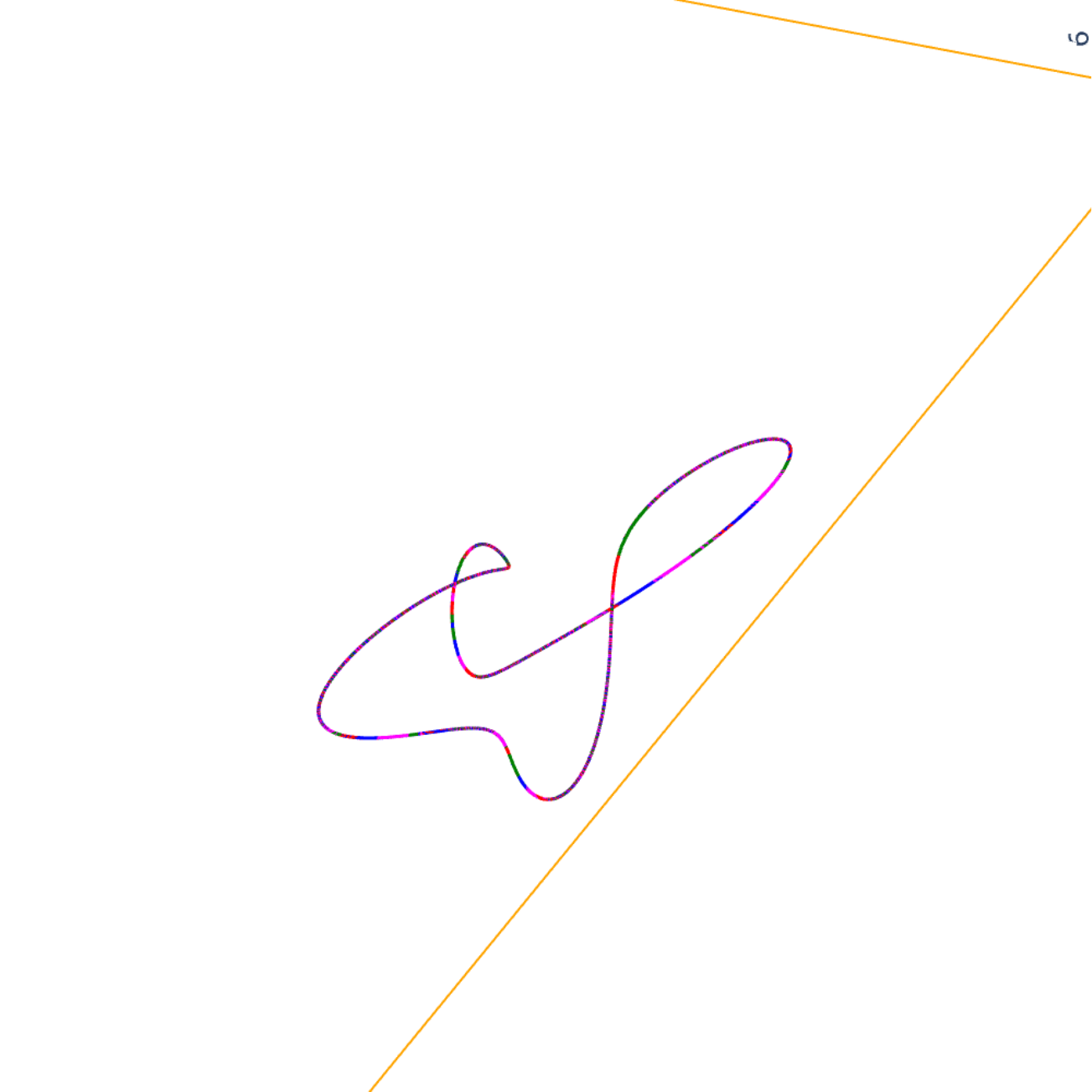}} \\ a) 
\end{minipage}
\begin{minipage}[h]{0.19\linewidth}
\center{\includegraphics[width=1\linewidth]{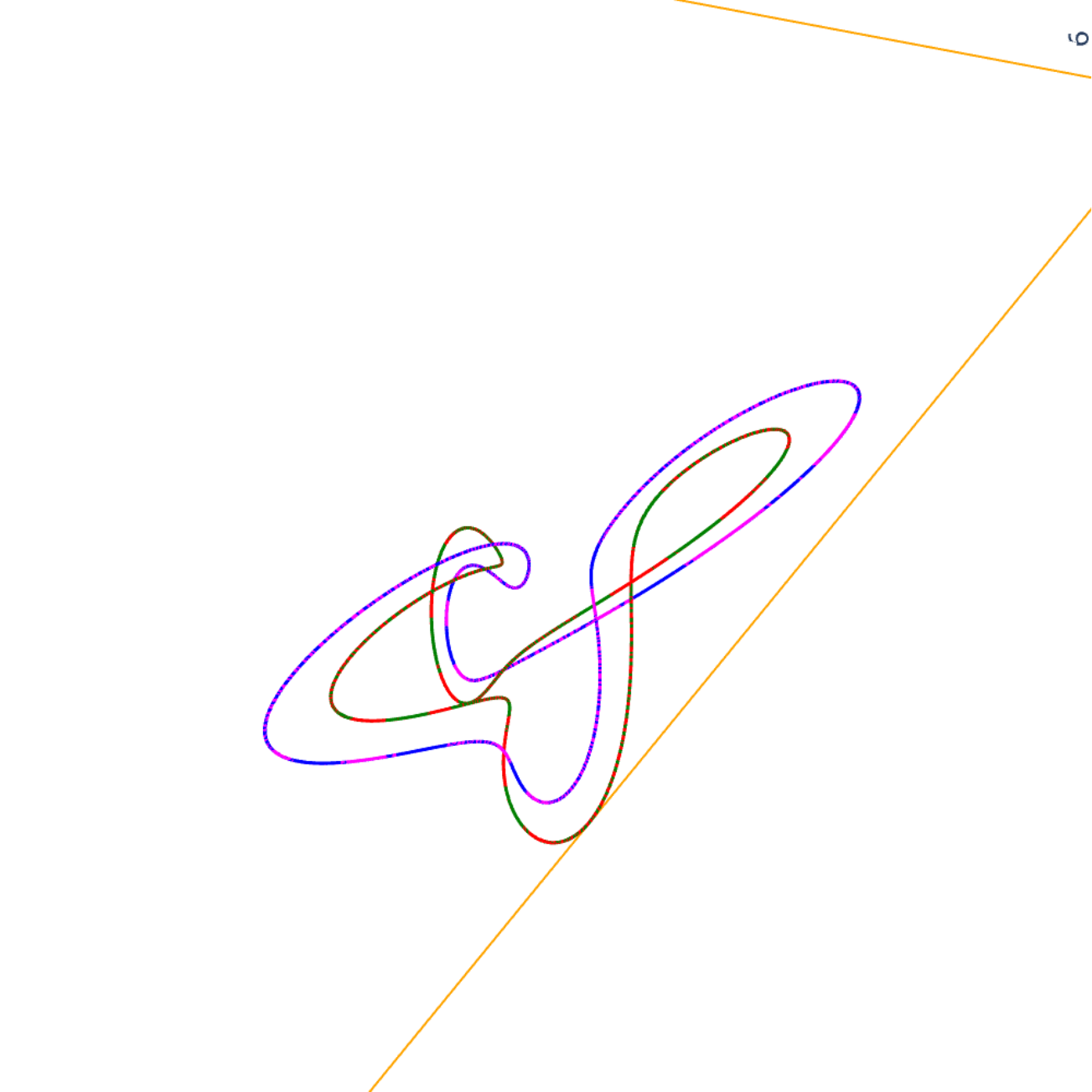}} \\b)
\end{minipage}
\begin{minipage}[h]{0.19\linewidth}
\center{\includegraphics[width=1\linewidth]{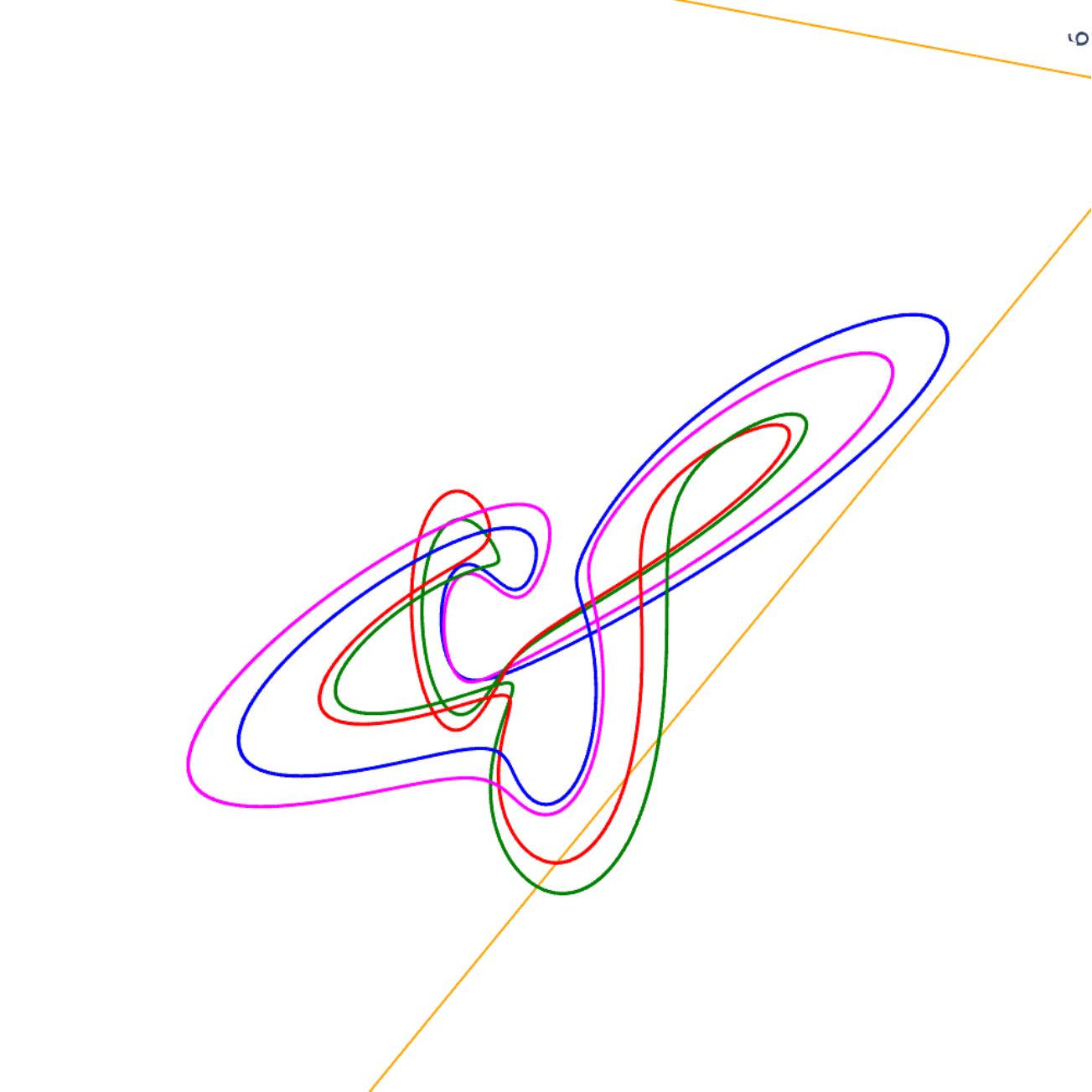}} \\ c) 
\end{minipage}
\begin{minipage}[h]{0.19\linewidth}
\center{\includegraphics[width=1\linewidth]{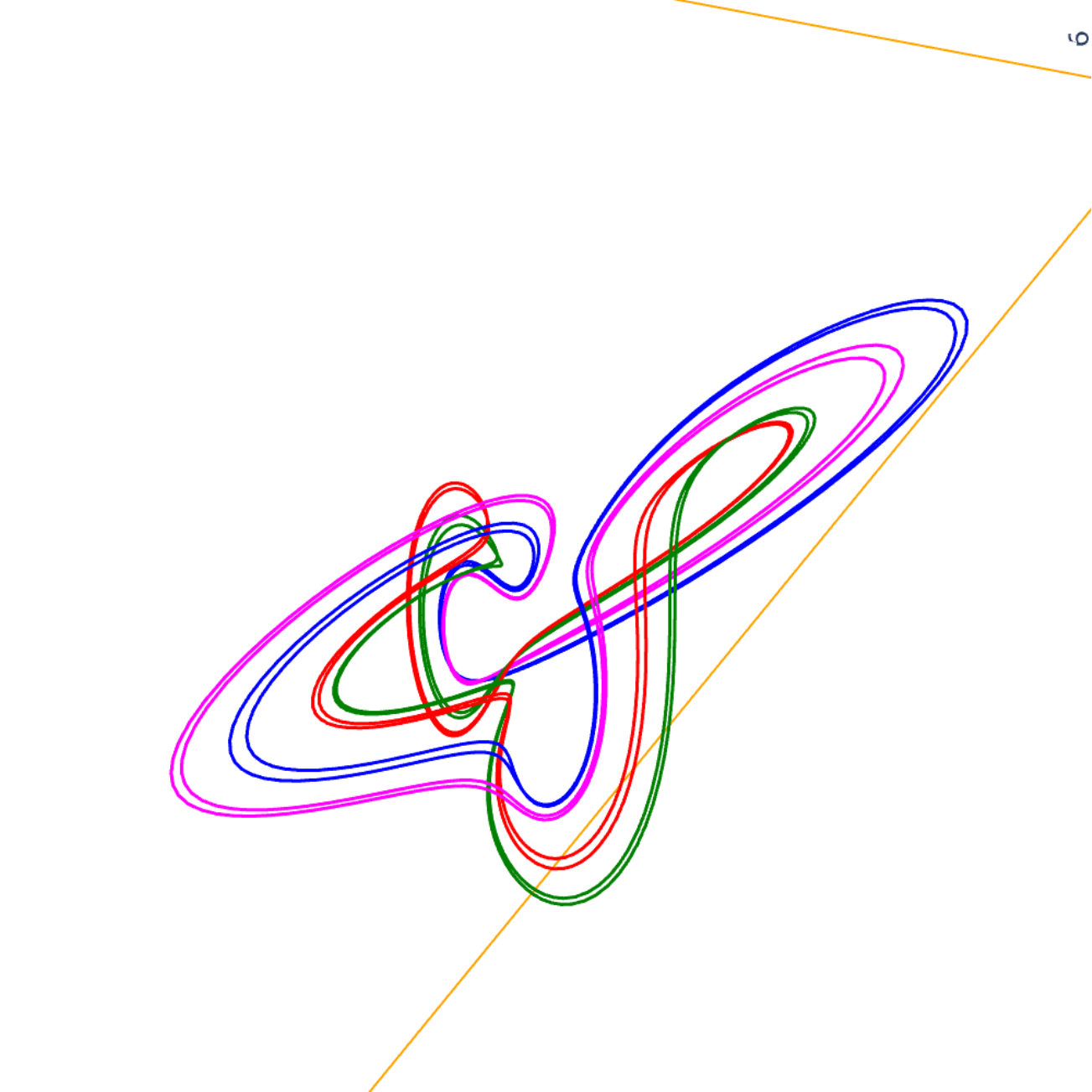}} \\ d) 
\end{minipage}
\begin{minipage}[h]{0.20\linewidth}
\center{\includegraphics[width=1\linewidth]{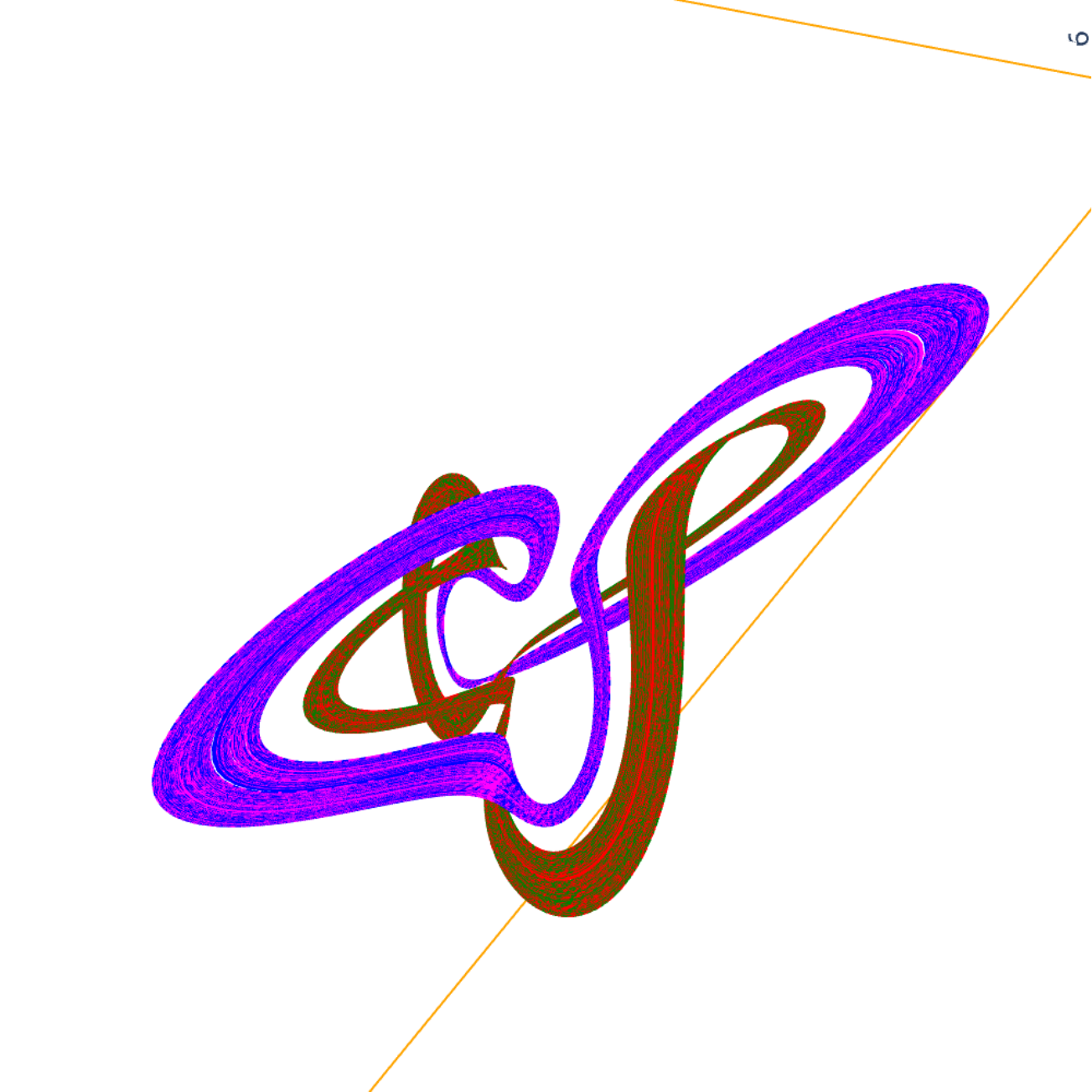}} \\ e) 
\end{minipage}
\caption{\label{fig:chaos_scenario_r}Attractors of the first scenario for different parameter values. The attractor $Attr$ is plotted with red color, while blue, green and magenta colors are used to depict its symmetric copies $T(Attr)$, $T^2(Attr)$ and $T^3(Attr)$ respectively: a) $r = 4$, b) $r = 3.1$, c) $r = 2.83$, d) $r = 2.79$, e) $r = 2.75,\, \Lambda \approx 0.0185$.}
\end{figure}

To illustrate the typical evolution of the attractors of the \eqref{eq:RedusedA4d} system, we turn to a one-parameter analysis.
In this scenario, the parameters $\alpha$ and $\beta$ are fixed and are assumed to be
$$\alpha = -2.911209192326542, \, \beta= -1.612684228842761.$$
Let us start describing the evolution of the attractor with a simple regime: the limit cycle found for $r = 4$ (Fig. \ref{fig:chaos_scenario_r}a).
This limit cycle is $T$-symmetric: that is, if we apply the mapping $T$ to it, then it maps to itself as a set.
As the parameter $r$ increases, this limit cycle disappears when the saddle-node bifurcation occurs.
At $r \approx 3.2537$, MATCONT seems to detect supercritical pitchfork bifurcation.
After this bifurcation, two stable symmetric limit cycles (Fig.~\ref{fig:chaos_scenario_r}b) appear in place of the old $T$-symmetric stable limit cycle.
These limit cycles turned out to be $T^2$-symmetric: that is, the attractor $Attr$ is such that $Attr \neq T(Attr)$, but $Attr = T^2(Attr)$.
At $r \approx 2.89$, the pitchfork bifurcation occurs again, further breaking the symmetry of the attractor.
For $r = 2.83$, four coexisting completely asymmetric limit cycles are demonstrated in Fig.~\ref{fig:chaos_scenario_r}c: each of these limit cycles $Attr$ satisfies $Attr \neq T(Attr)$ and $Attr \neq T^2(Attr)$.
A further decrease in $r$ is accompanied by a cascade of period doubling bifurcations (the first doubling is shown in Fig.~\ref{fig:chaos_scenario_r}d).
For $r = 2.75$, the system already has a $T^2$-symmetric chaotic attractor (Fig.~\ref{fig:chaos_scenario_r}e) with the highest Lyapunov exponent $\Lambda \approx 0.0185$.
Partial restoration of the attractor symmetry may have occurred at the reverse supercritical pitchfork bifurcation in some periodicity window.
\subsection{Emergence of chaos through a sequence of period-doubling bifurcations and subsequent merging of symmetric chaotic attractors}

Let us demonstrate the second scenario for the emergence and evolution of chaotic attractors. We will now change the parameter $\alpha$ for fixed values of $\beta = -1.612684228842761$ and $r = 1$.
\begin{figure}[!thb]
\centering
\begin{minipage}[h]{0.20\linewidth}
\center{\includegraphics[width=1\linewidth]{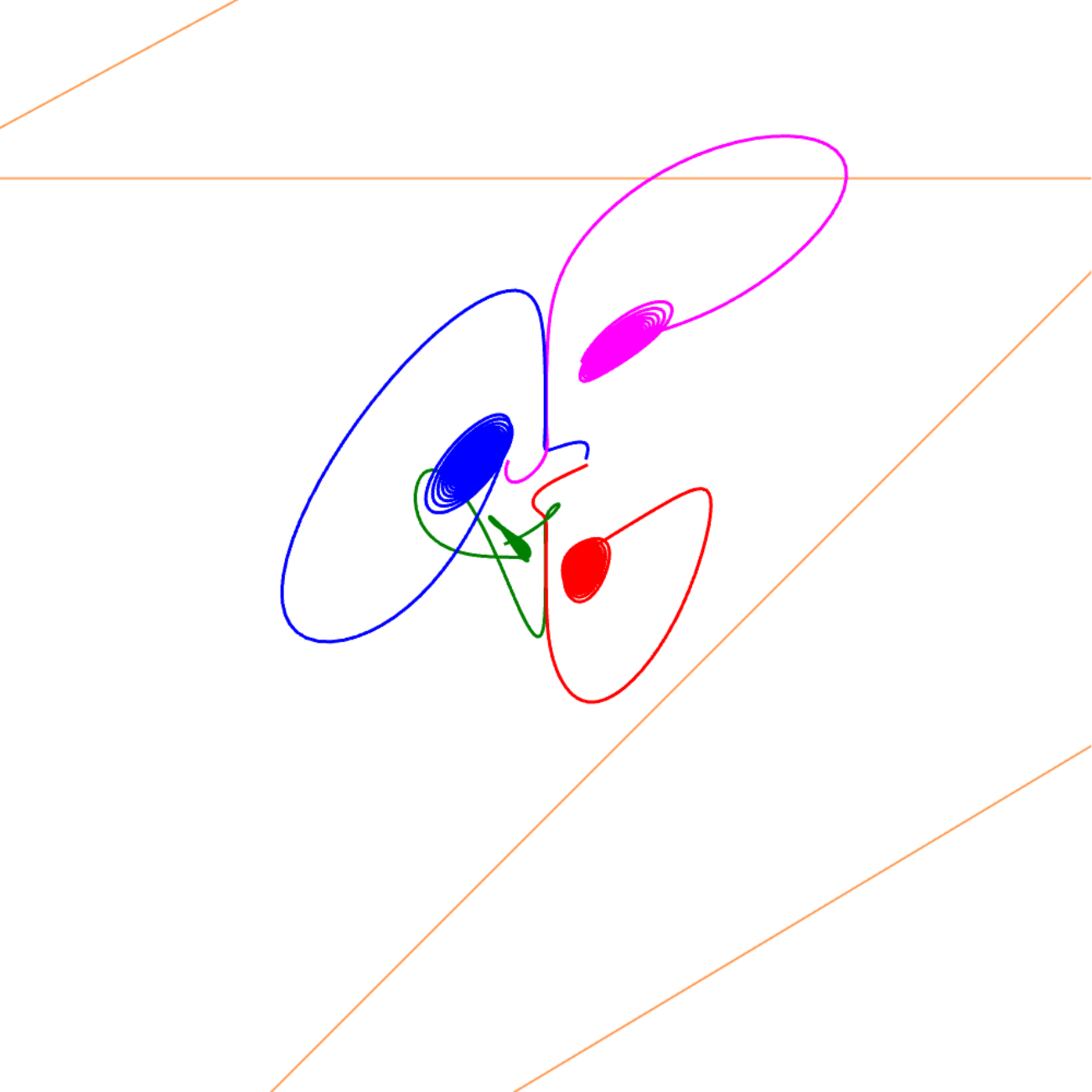}} \\ a)
\end{minipage}
\begin{minipage}[h]{0.20\linewidth}
\center{\includegraphics[width=1\linewidth]{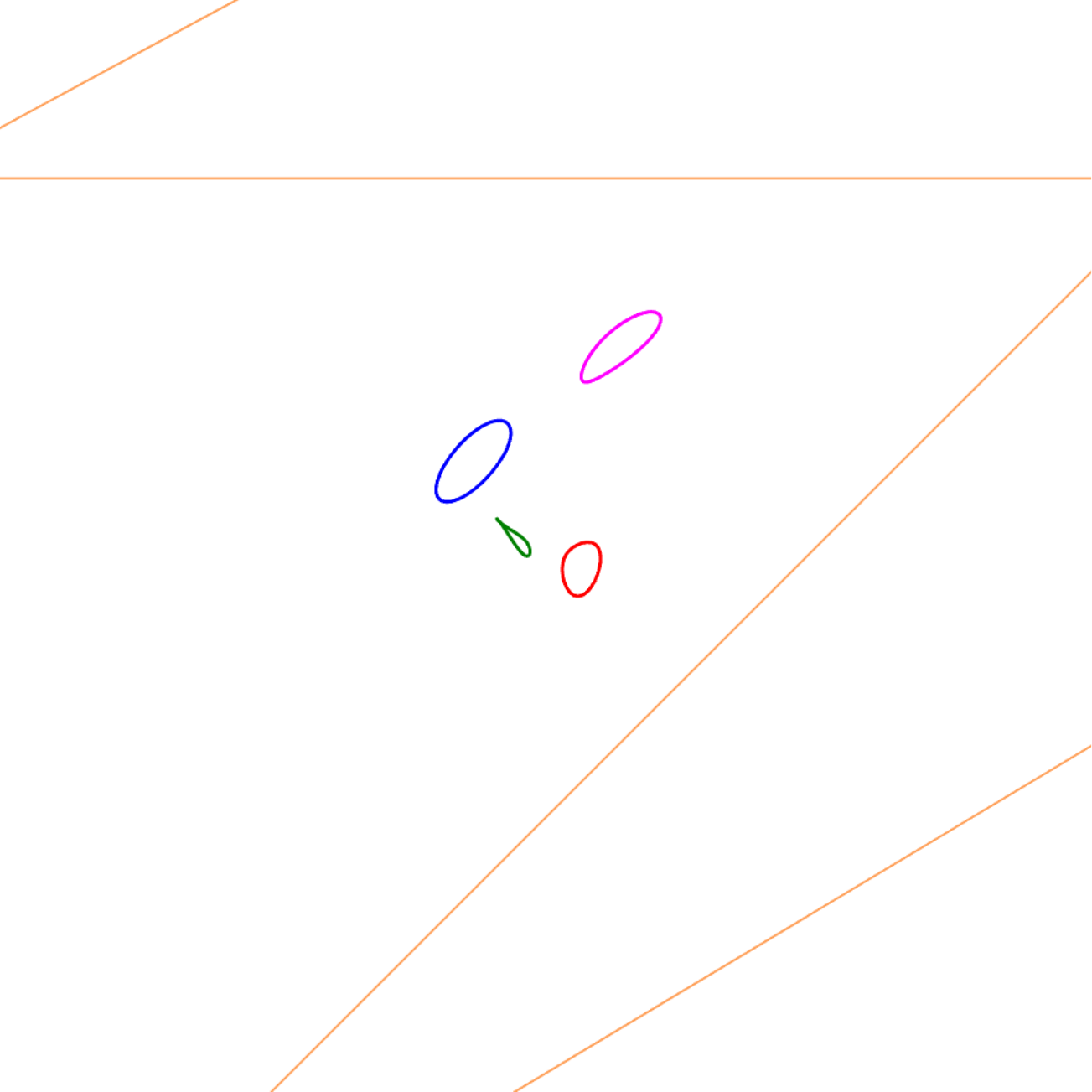}} \\ b)
\end{minipage}
\begin{minipage}[h]{0.20\linewidth}
\center{\includegraphics[width=1\linewidth]{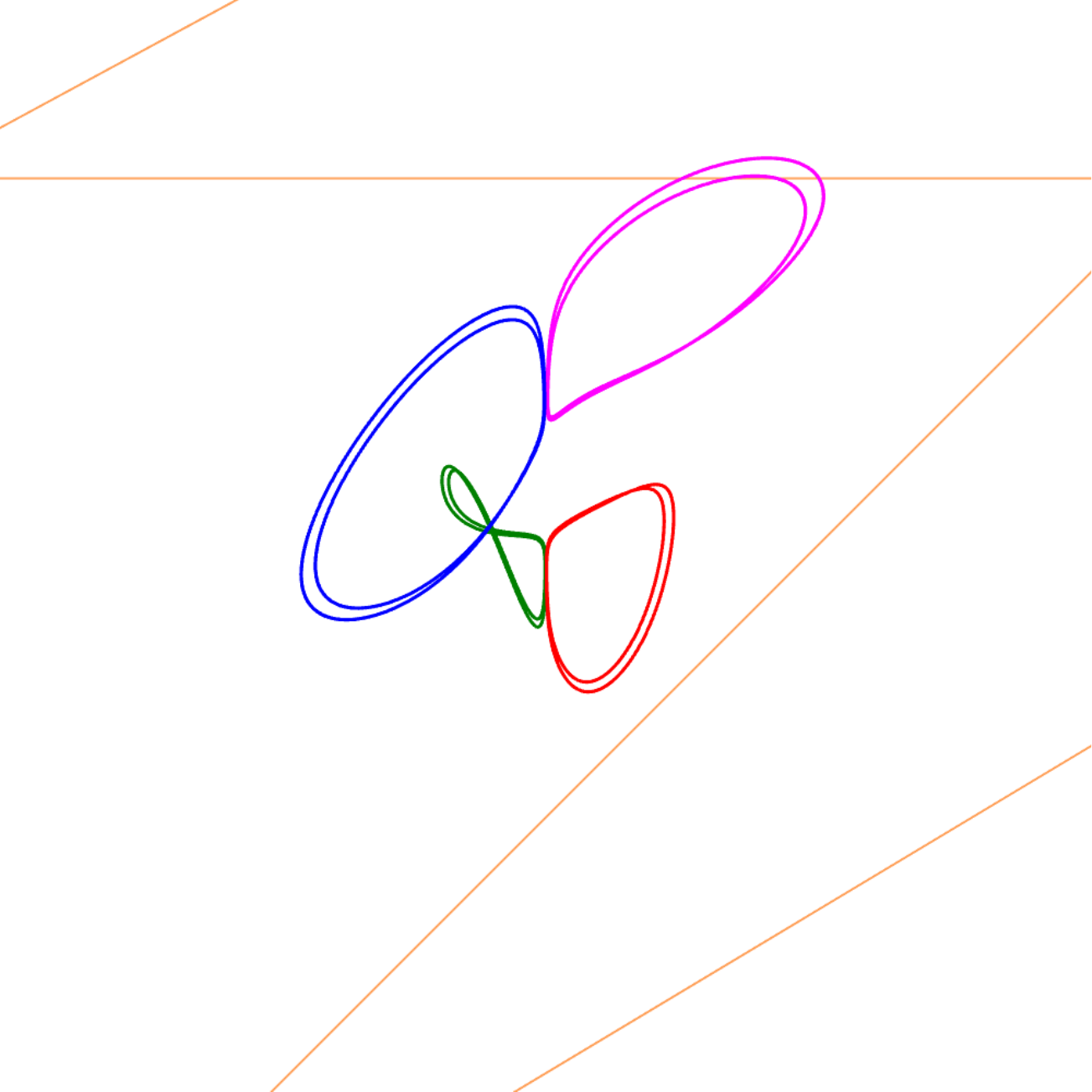}} \\ c)
\end{minipage}
\begin{minipage}[h]{0.20\linewidth}
\center{\includegraphics[width=1\linewidth]{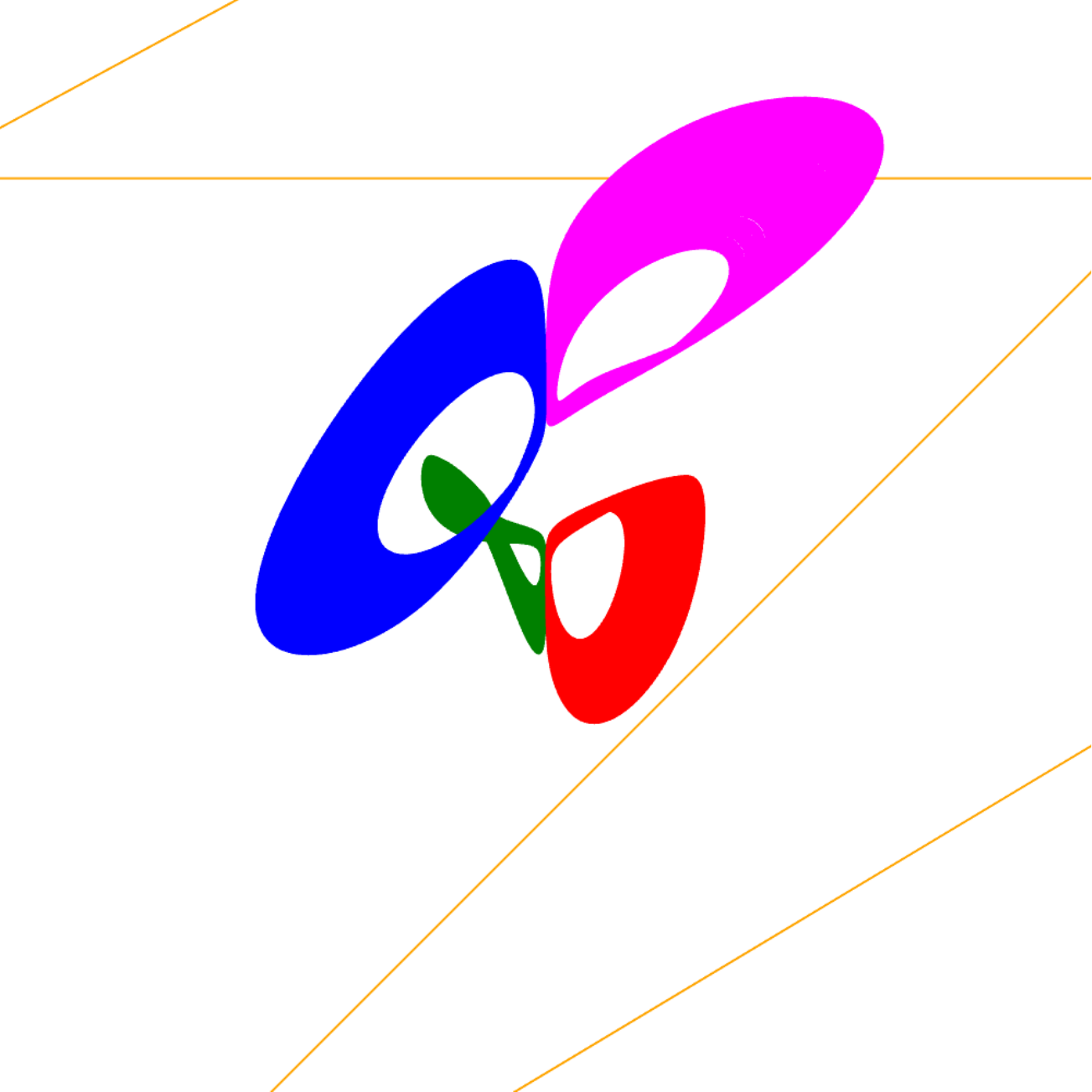}} \\ d)
\end{minipage}
\vfill
\centering
\begin{minipage}[h]{0.20\linewidth}
\center{\includegraphics[width=1\linewidth]{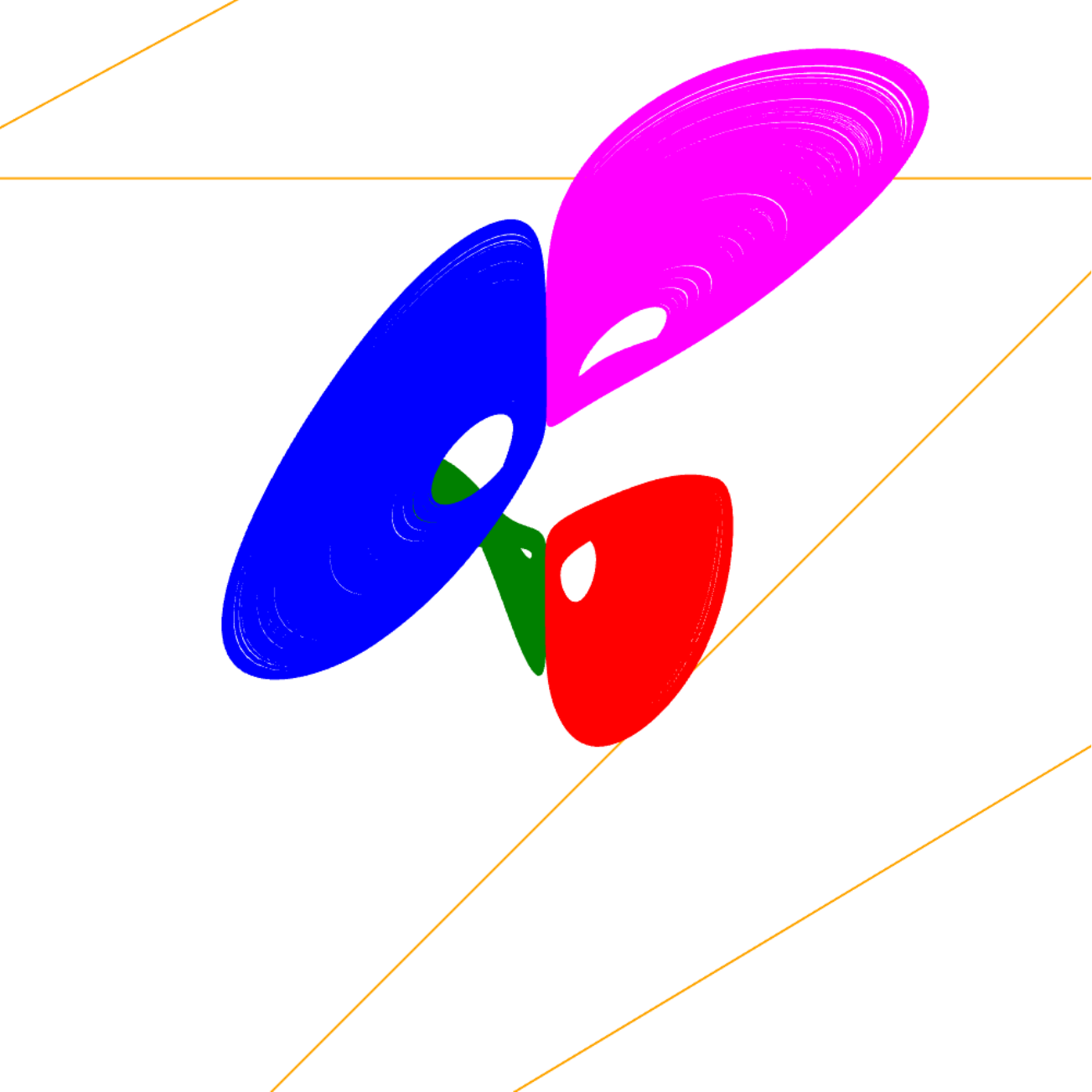}} \\ e)
\end{minipage}
\begin{minipage}[h]{0.20\linewidth}
\center{\includegraphics[width=1\linewidth]{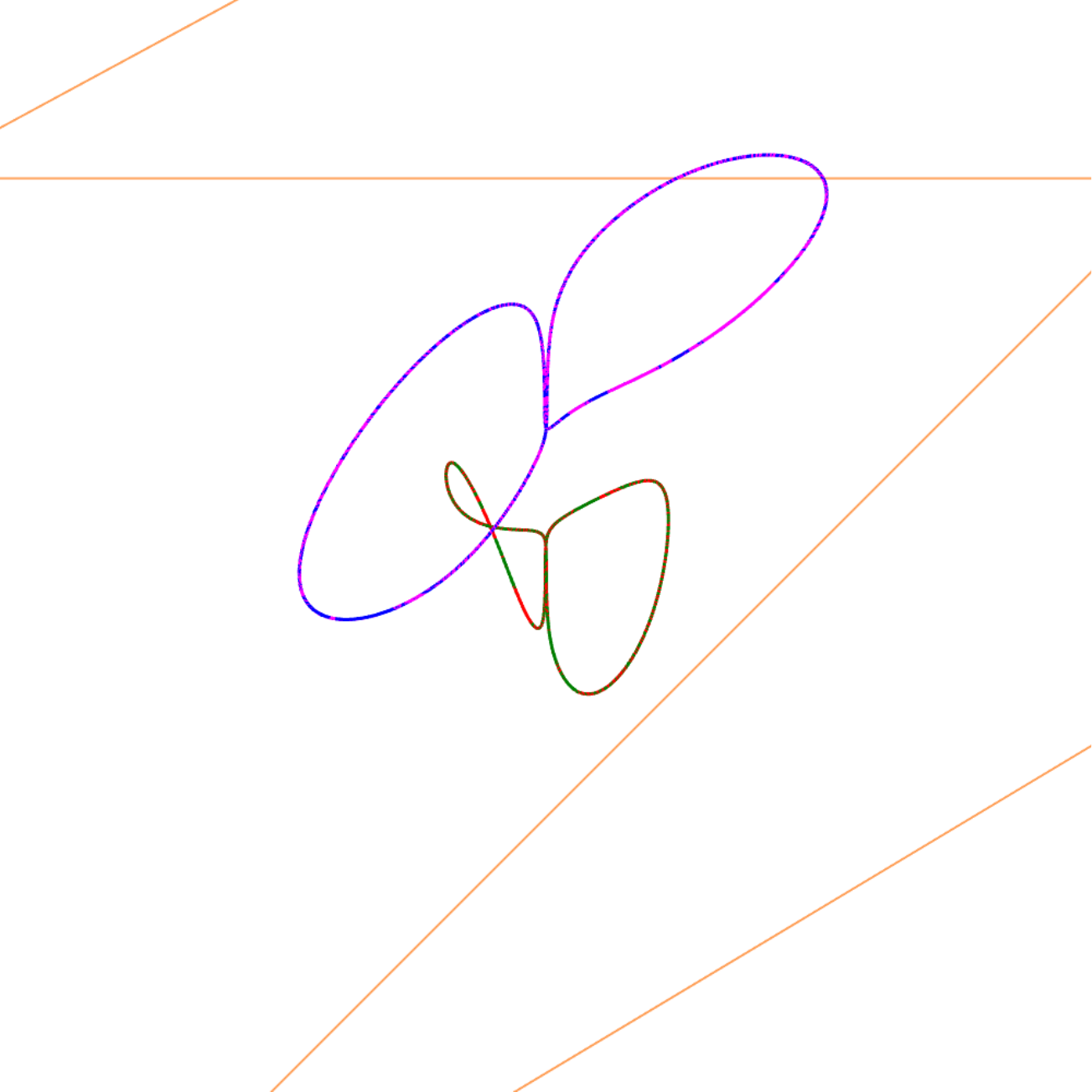}} \\ f)
\end{minipage}
\begin{minipage}[h]{0.20\linewidth}
\center{\includegraphics[width=1\linewidth]{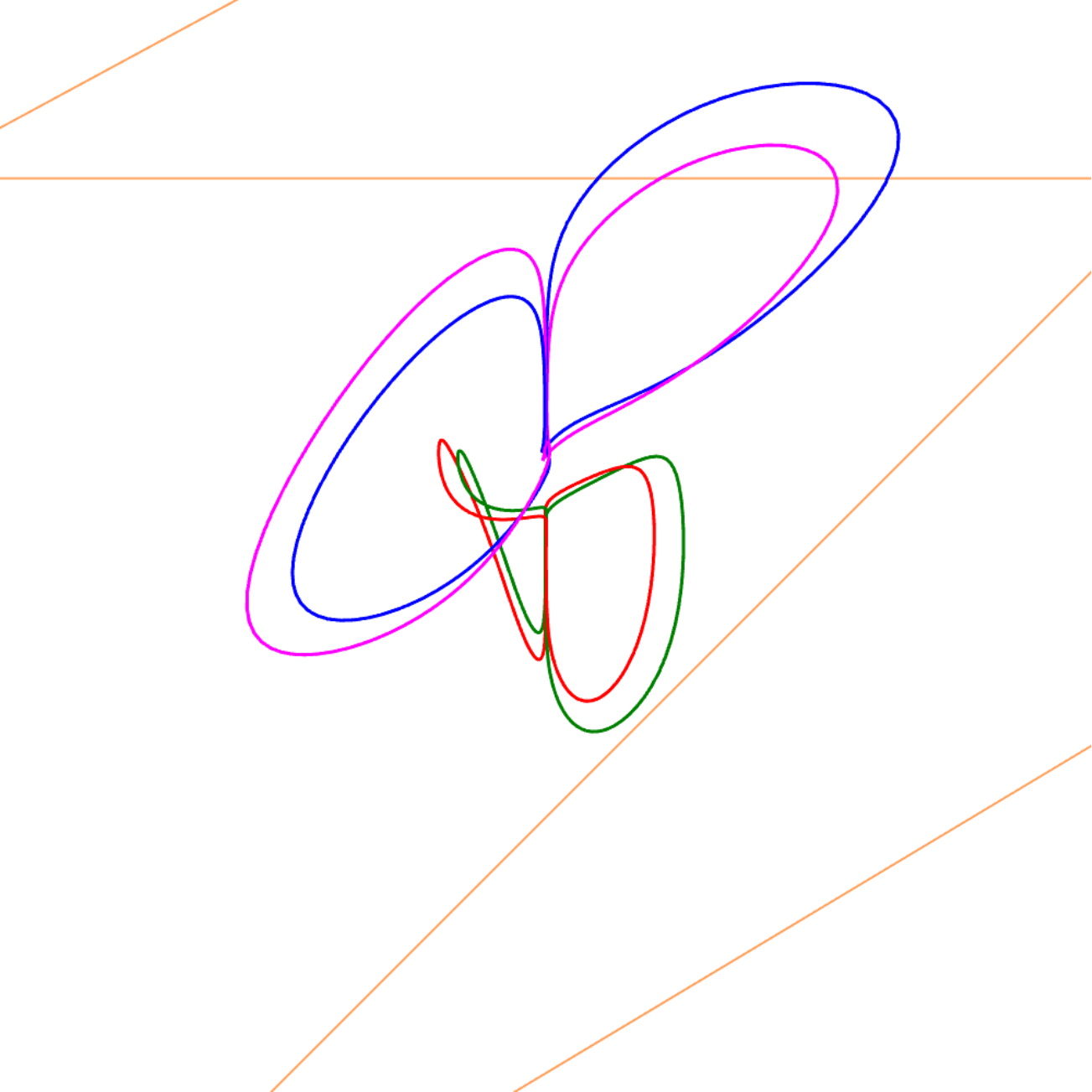}} \\ g)
\end{minipage}
\begin{minipage}[h]{0.20\linewidth}
\center{\includegraphics[width=1\linewidth]{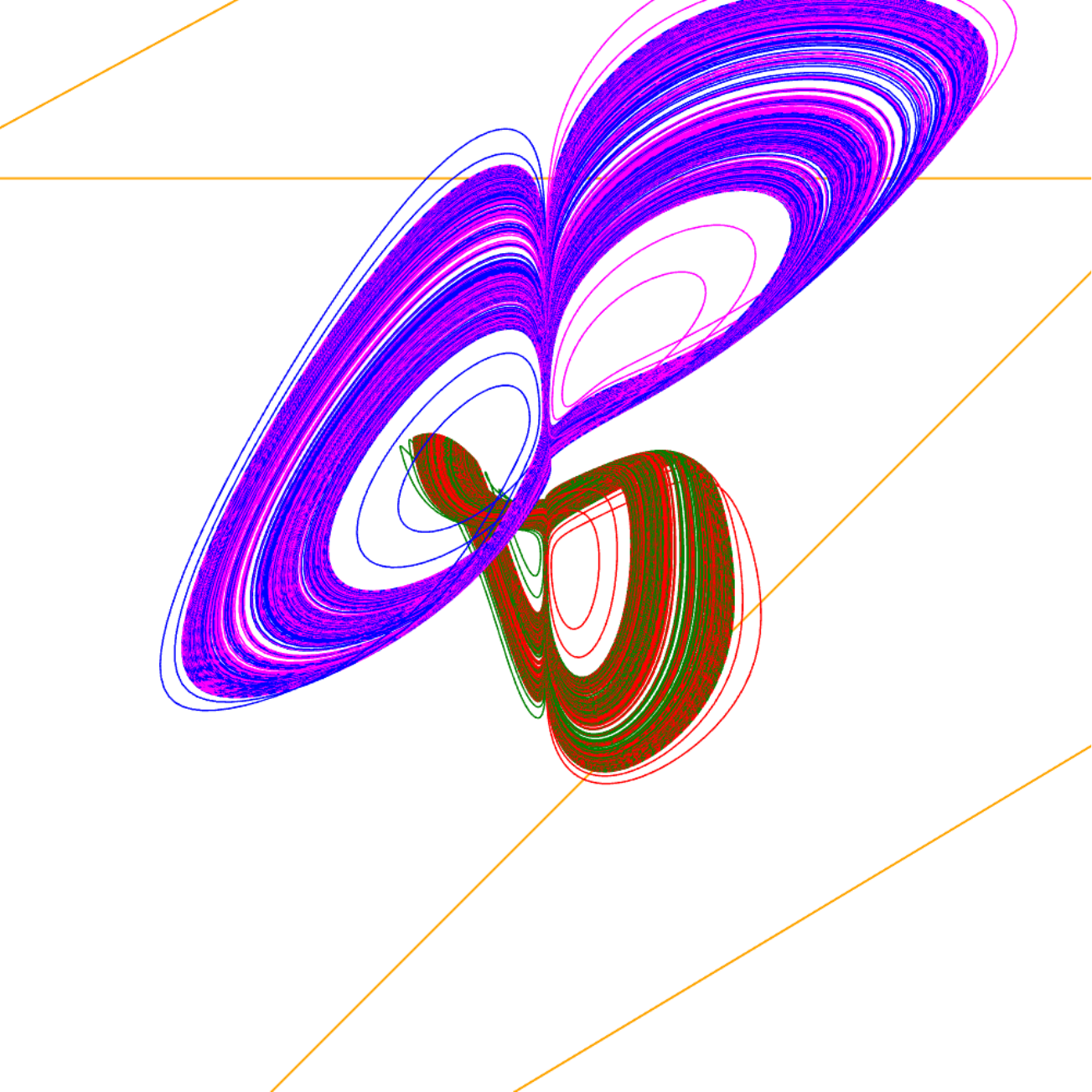}} \\ h)
\end{minipage}
\vfill
\begin{minipage}[h]{0.20\linewidth}
\center{\includegraphics[width=1\linewidth]{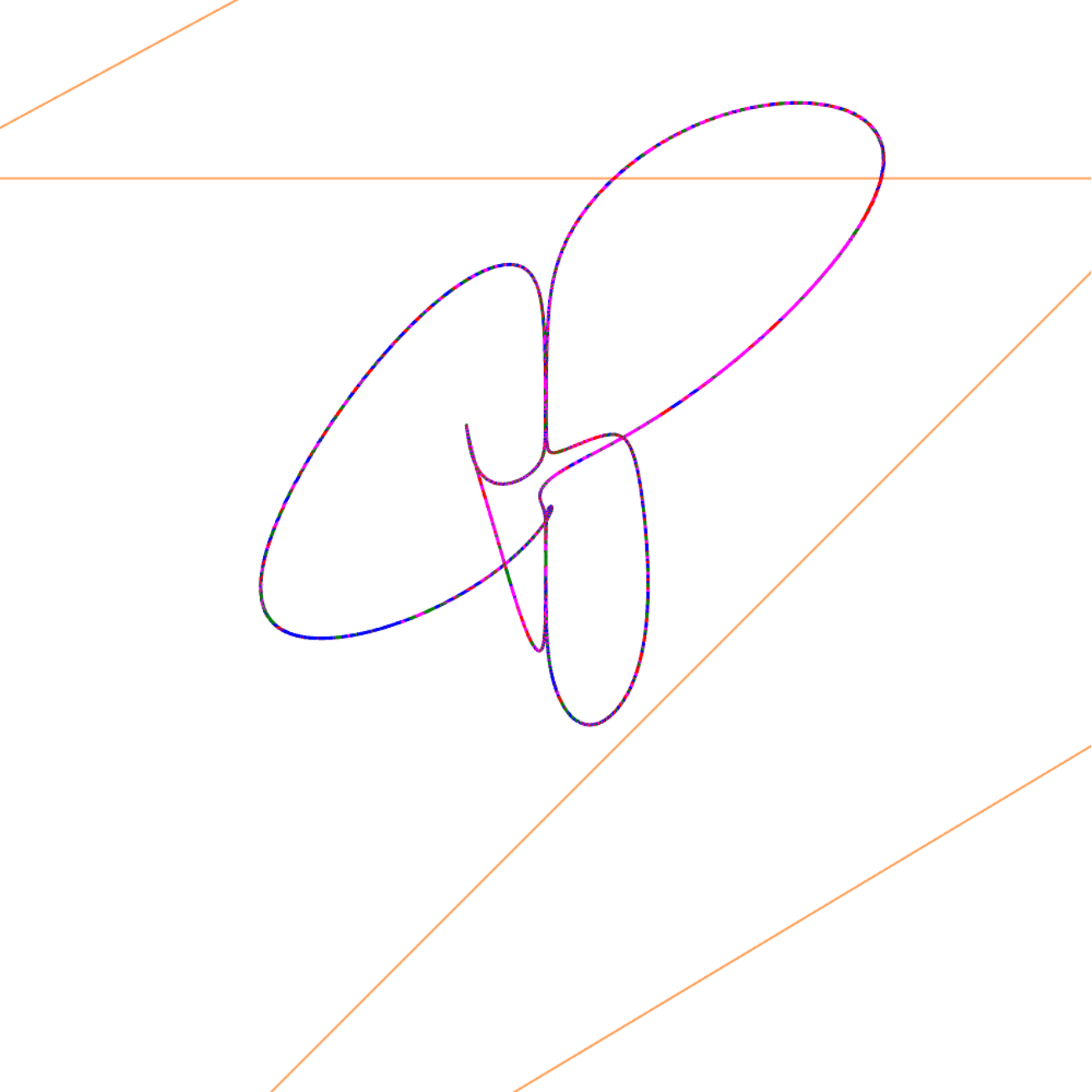}} \\ i)
\end{minipage}
\begin{minipage}[h]{0.20\linewidth}
\center{\includegraphics[width=1\linewidth]{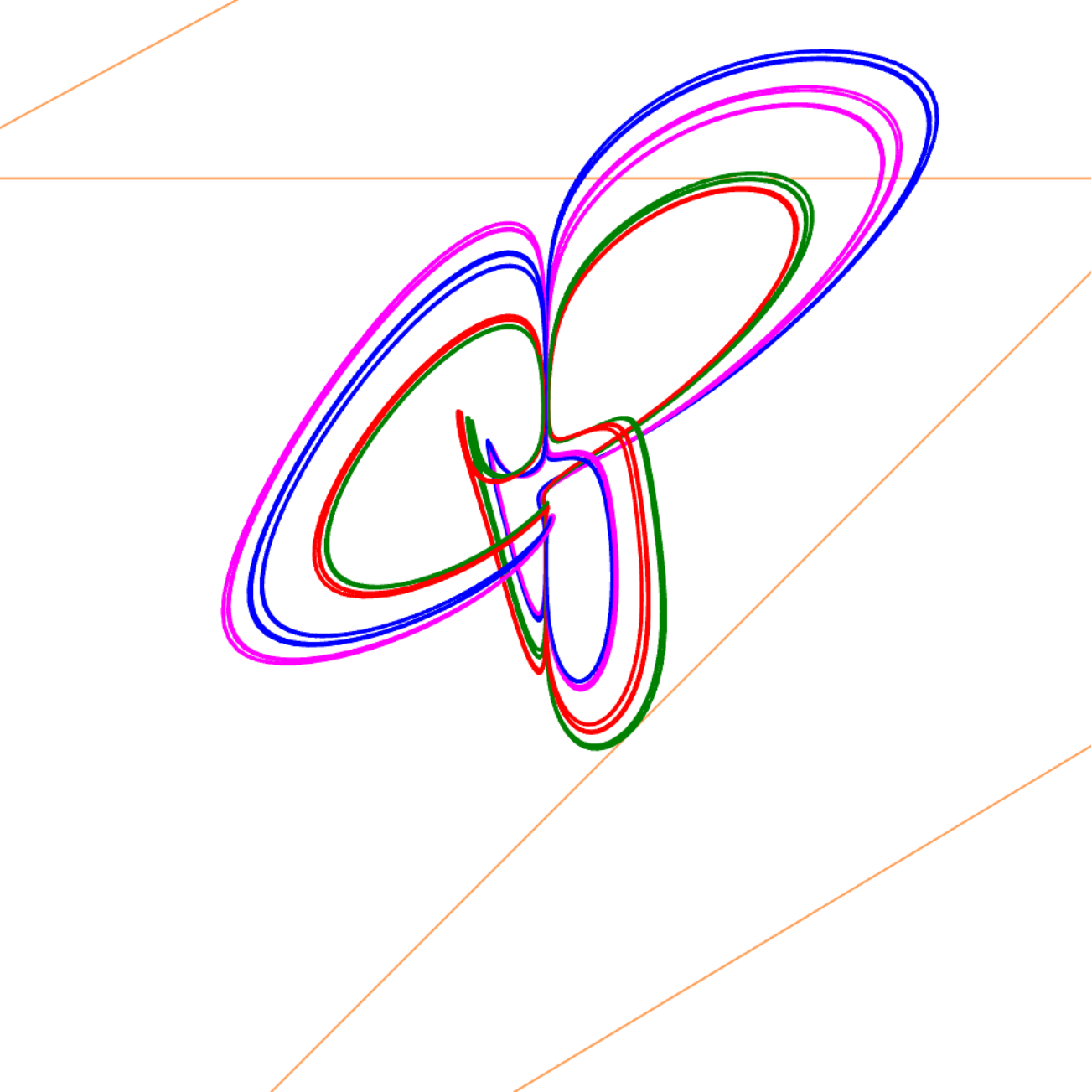}} \\ j)
\end{minipage}
\begin{minipage}[h]{0.20\linewidth}
\center{\includegraphics[width=1\linewidth]{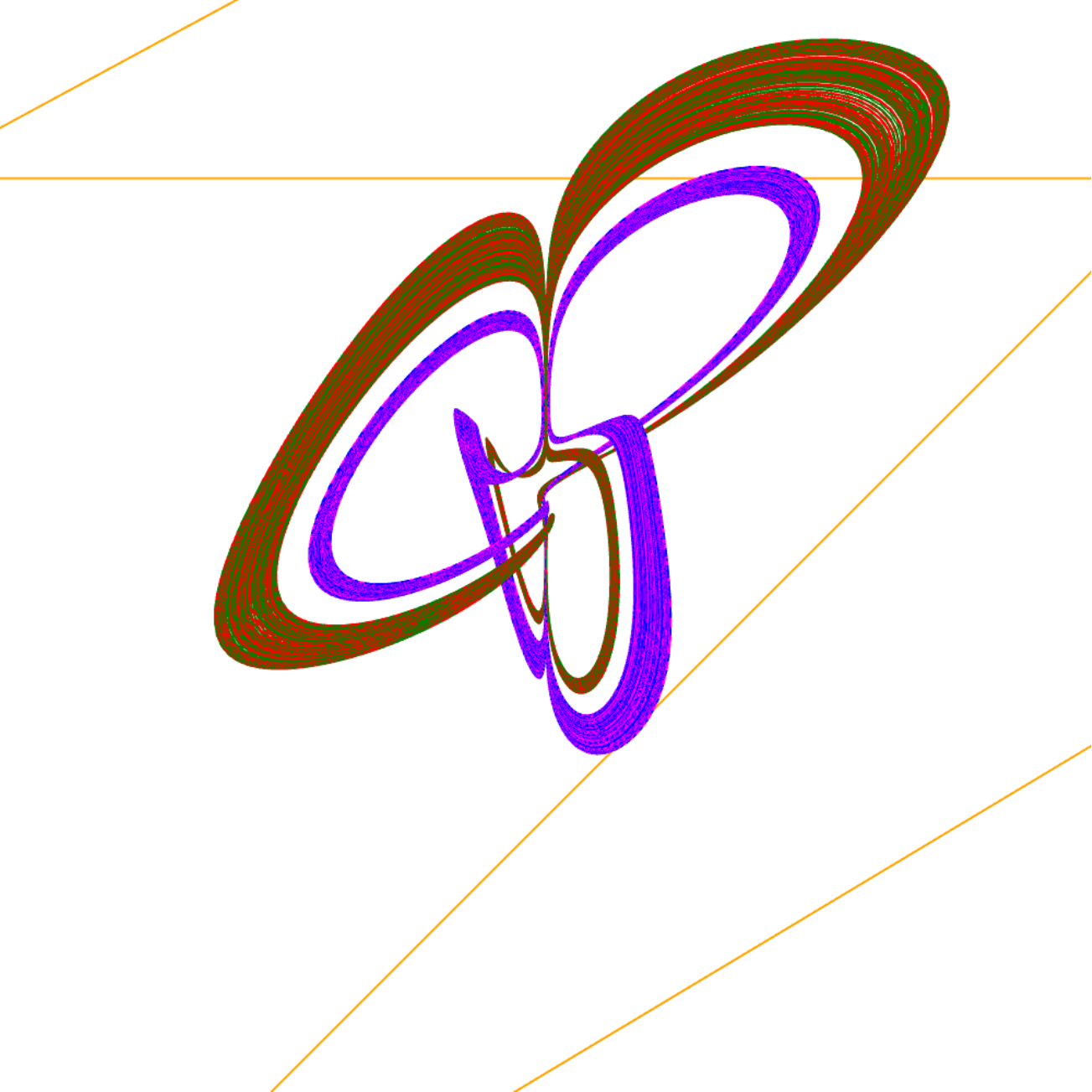}} \\ k)
\end{minipage}
\centering
\begin{minipage}[h]{0.20\linewidth}
\center{\includegraphics[width=1\linewidth]{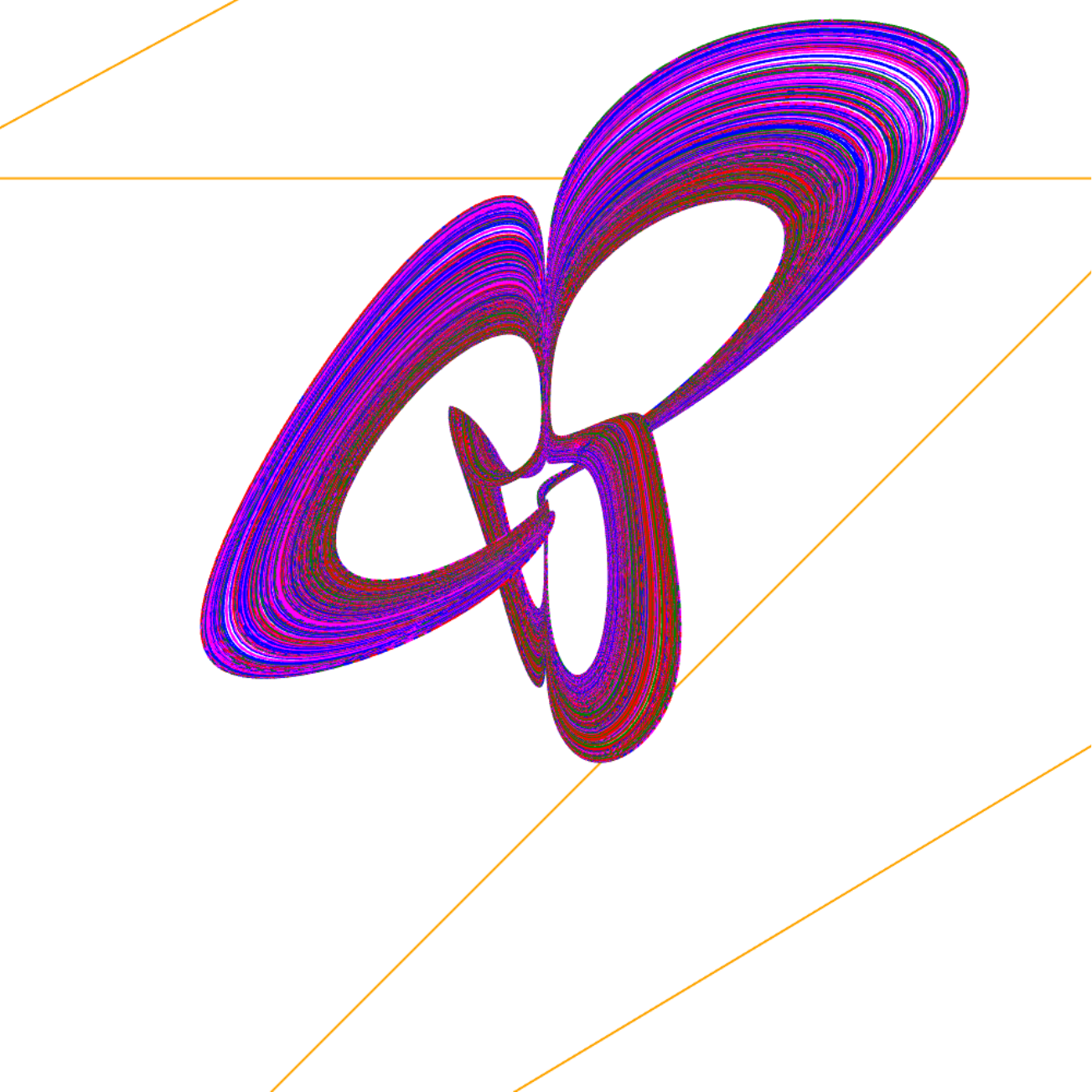}} \\ l)
\end{minipage}
\caption{\label{fig:chaos_scenario_alpha}Attractors of the second scenario at different parameter values. The largest Lyapunov exponent $\Lambda$ is written when the attractor is chaotic:
a) $\alpha = -2.32$; b) $\alpha = -2.37$; c) $\alpha = -2.47$; d) $\alpha = -2.484, \Lambda \approx 0.0195$;
e) $\alpha = -2.487, \Lambda \approx 0.0224$; f) $\alpha = -2.515$; g)  $\alpha = -2.64$; h) $\alpha = -2.67, \Lambda \approx 0.014$;
i) $\alpha = -2.799$; j)  $\alpha = -2.835$;  k) $\alpha = -2.837, \Lambda \approx 0.07$; l) $\alpha = -2.84, \Lambda \approx 0.0087$.}
\end{figure}
For $\alpha = -2.32$, there are 4 stable equilibrium states inside the canonical invariant region (Fig.~\ref{fig:chaos_scenario_alpha}a): all eigenvalues of the Jacobi matrix are to the left of the imaginary axis on the complex plane, and the closest to the imaginary axis is a pair of complex conjugate numbers.
As $\alpha$ decreases, Andronov-Hopf bifurcation occurs, after which 4 asymmetric stable limit cycles are observed in the system (Fig. \ref{fig:chaos_scenario_alpha}b).
As the parameter decreases further from $-2.37$ to $-2.47$, the limit cycle grows in size and multiple period doubling bifurcations occur (Fig.~\ref{fig:chaos_scenario_alpha}c shows the first doubling).
The doubling cascade gives rise to a chaotic attractor at $\alpha = -2.484$ (Fig.~\ref{fig:chaos_scenario_alpha}d).
Then the attractor further increases in size (Fig.~\ref{fig:chaos_scenario_alpha}e), and the "hole" in it becomes smaller, which may indicate the subsequent formation of Shilnikov's homoclinic attractor \cite{BGGKS2022}.
The symmetry properties of the attractor has not changed over the entire range of the parameter: none of the four attractors coincided with their images under the action of any of the nonzero powers of the mapping $T$.
A further change in the parameter leads to a change in the symmetry type of the attractor: for $\alpha = -2.515$ attractors (a pair of limit cycles, see Fig. \ref{fig:chaos_scenario_alpha}f) are already $T^2$-symmetric, and for $\alpha = -2.64$ there are four asymmetric attractors again (Fig. \ref{fig:chaos_scenario_alpha}g).
The next sequence of period doubling bifurcations leads to the appearance of a chaotic attractor and subsequent partial restoration of symmetry ($T^2$-symmetric chaotic attractor in Fig. \ref{fig:chaos_scenario_alpha}h), after which the attractor is a $T$-symmetric limit cycle (Fig. \ref{fig:chaos_scenario_alpha}i).
The further evolution of the attractor in a sense repeats the previous stages: gradual destruction of symmetry to asymmetric attractors and a sequence of period doubling bifurcations (Fig. \ref{fig:chaos_scenario_alpha}j), chaotization and partial restoration of symmetry (Fig. \ref{fig:chaos_scenario_alpha}k) ending with the formation of a $T$-symmetric chaotic attractor (Fig. \ref{fig:chaos_scenario_alpha}l).

\section{Conclusion}

In this paper, we have established for the first time the existence of chaotic attractors in a system of four globally coupled phase oscillators with a biharmonic coupling function.
Similar to the attractors studied in \cite{BAR16, GO18}, the examples we found seem to be closely related to this system's symmetry group and the heteroclinic cycles that arise due to it.
This symmetry affects both the shape of the heteroclinic cycles themselves and, apparently, the shape of the chaotic attractors that emerge nearby.
Other attractors that we have found can emerge from the regular cascade of period doublings of limit cycles -- a scenario that is common for generic dissipative systems and not related to symmetry properties of system.
However, the presence of symmetry affects the further evolution of such attractors, resulting in symmetry-breaking or symmetry-restoring bifurcations \cite{ChGo1988}.
Fig.~\ref{fig:HetMap} also highlights a peculiar proximity of heteroclinic cycles built on different sets of connections $\sigma_{\rm s} \rightarrow \sigma_{\rm sf} \rightarrow T^k(\sigma_{\rm s})$.
Intuitively and non-rigorously, this probably can be explained as follows: by changing the parameters, we can destroy the heteroclinic cycle $\sigma_{\rm s} \rightarrow \sigma_{\rm sf} \rightarrow T(\sigma_{\rm s})$ in such a way that the separatrix $\gamma_{\rm sf}$ passes near the equilibrium state $T(\sigma_{\rm s})$, but approaches the equilibrium state $T^2(\sigma_{\rm s})$ along its stable manifold.
A more detailed investigation of the relationship between the symmetries of attractors and heteroclinic cycles is not the subject of this work, but will be a part of our future research.
Finally, we note that the proposed approach for the search for chaotic attractors can be applied to any system of four globally coupled phase oscillators of Kuramoto-type, regardless of the coupling function.
Moreover, it can also be applied to systems with non-pairwise interactions if the equations depend on the phase differences in the same fashion as in Kuramoto-type system \eqref{eq:KurTypeGlobalIdentical}.
This approach can provide another way to explore systems for which there are currently no numerical evidences of chaotic dynamics, for example, for the phase reduction of four oscillators in the mean-field complex Ginzburg-Landau equation \cite{LeonPazo2019}.

\section*{Acknowledgements}
The results in Section 2  were supported by Ministry of Science and Higher Education of Russian Federation, contract
0729-2020-0036. The results in Section 3 were supported by RSF grant 22-12-00348.

\bibliographystyle{alpha}
\bibliography{refer}

\newcommand{\etalchar}[1]{$^{#1}$}
\begin{thebibliography}{AOWT07}

\bibitem[AB15]{AshBur2015}
Peter Ashwin and Oleksandr Burylko.
\newblock {Weak chimeras in minimal networks of coupled phase oscillators}.
\newblock {\em Chaos: An Interdisciplinary Journal of Nonlinear Science},
  25(1):013106, 2015.

\bibitem[ABB16]{ABB16}
Peter Ashwin, Christian Bick, and Oleksandr Burylko.
\newblock Identical phase oscillator networks: Bifurcations, symmetry and
  reversibility for generalized coupling.
\newblock {\em Frontiers in Applied Mathematics and Statistics}, 2:7, June
  2016.

\bibitem[ABR22]{AshBickRodr2022}
Peter Ashwin, Christian Bick, and Ana Rodrigues.
\newblock From symmetric networks to heteroclinic dynamics and chaos in coupled
  phase oscillators with higher-order interactions.
\newblock In Federico Battiston and Giovanni Petri, editors, {\em Understanding
  Complex Systems}, pages 197--216. Springer International Publishing, 2022.

\bibitem[ABV{\etalchar{+}}05]{AceBonVicRitSpig2005}
Juan~A. Acebr{\'{o}}n, L.~L. Bonilla, Conrad~J.P{\'{e}}rez Vicente, F{\'{e}}lix
  Ritort, and Renato Spigler.
\newblock {The Kuramoto model: A simple paradigm for synchronization
  phenomena}.
\newblock {\em Reviews of Modern Physics}, 77(1):137--185, 2005.

\bibitem[AOWT07]{AOWT07}
Peter Ashwin, Gábor Orosz, John Wordsworth, and Stuart Townley.
\newblock Dynamics on networks of cluster states for globally coupled phase
  oscillators.
\newblock {\em SIAM Journal on Applied Dynamical Systems}, 6(4):728--758, 2007.

\bibitem[AR16]{AshRodr2016}
Peter Ashwin and Ana Rodrigues.
\newblock {Hopf normal form with $S_N$ symmetry and reduction to systems of
  nonlinearly coupled phase oscillators}.
\newblock {\em Physica D: Nonlinear Phenomena}, 325:14--24, 2016.

\bibitem[AS92]{AshSwift1992}
P.~Ashwin and J.~W. Swift.
\newblock {The dynamics of $n$ weakly coupled identical oscillators}.
\newblock {\em Journal of Nonlinear Science}, 2(1):69--108, 1992.

\bibitem[AS04]{AbrStrog2004}
Daniel~M Abrams and Steven~H Strogatz.
\newblock Chimera states for coupled oscillators.
\newblock {\em Physical review letters}, 93(17):174102, 2004.

\bibitem[BA16]{BickAsh2016}
Christian Bick and Peter Ashwin.
\newblock Chaotic weak chimeras and their persistence in coupled populations of
  phase oscillators.
\newblock {\em Nonlinearity}, 29(5):1468, mar 2016.

\bibitem[BAR16]{BAR16}
Christian Bick, Peter Ashwin, and Ana Rodrigues.
\newblock Chaos in generically coupled phase oscillator networks with
  nonpairwise interactions.
\newblock {\em Chaos: An Interdisciplinary Journal of Nonlinear Science},
  26(9):094814, September 2016.

\bibitem[BGG{\etalchar{+}}22]{BGGKS2022}
Yu~V Bakhanova, SV~Gonchenko, AS~Gonchenko, AO~Kazakov, and EA~Samylina.
\newblock {On Shilnikov attractors of three-dimensional flows and maps}.
\newblock {\em Journal of Difference Equations and Applications}, pages 1--18,
  2022.

\bibitem[BGHS21]{BickGroHarScha2021}
Christian Bick, Elizabeth Gross, Heather~A. Harrington, and Michael~T. Schaub.
\newblock {What are higher-order networks?}
\newblock {\em arXiv}, 2021.

\bibitem[Bic17]{Bick2017}
Christian Bick.
\newblock Isotropy of angular frequencies and weak chimeras with broken
  symmetry.
\newblock {\em Journal of nonlinear science}, 27(2):605--626, 2017.

\bibitem[Bic18]{Bick2018}
Christian Bick.
\newblock Heteroclinic switching between chimeras.
\newblock {\em Phys. Rev. E}, 97:050201, May 2018.

\bibitem[Bic19]{Bick2019}
Christian Bick.
\newblock {Heteroclinic dynamics of localized frequency synchrony: heteroclinic
  cycles for small populations}.
\newblock {\em Journal of Nonlinear Science}, 29(6):2547--2570, 2019.

\bibitem[BL19]{BickLohse2019}
Christian Bick and Alexander Lohse.
\newblock {Heteroclinic dynamics of localized frequency synchrony: Stability of
  heteroclinic cycles and networks}.
\newblock {\em Journal of Nonlinear Science}, 29(6):2571--2600, 2019.

\bibitem[BMB22]{BurMarBick2022}
Oleksandr Burylko, Erik~A. Martens, and Christian Bick.
\newblock {Symmetry breaking yields chimeras in two small populations of
  Kuramoto-type oscillators}.
\newblock {\em Chaos: An Interdisciplinary Journal of Nonlinear Science},
  32(9):093109, sep 2022.

\bibitem[BPM18]{BiPaMa2018}
Christian Bick, Mark~J Panaggio, and Erik~A Martens.
\newblock {Chaos in Kuramoto oscillator networks}.
\newblock {\em Chaos: An Interdisciplinary Journal of Nonlinear Science},
  28(7):071102, 2018.

\bibitem[BTP{\etalchar{+}}11]{BickTimPaulRathAsh2011}
Christian Bick, Marc Timme, Danilo Paulikat, Dirk Rathlev, and Peter Ashwin.
\newblock {Chaos in Symmetric Phase Oscillator Networks}.
\newblock {\em Phys. Rev. Lett.}, 107:244101, Dec 2011.

\bibitem[Bur20]{burylko2020collective}
OA~Burylko.
\newblock {Collective dynamics and bifurcations in symmetric networks of phase
  oscillators. I}.
\newblock {\em Journal of Mathematical Sciences}, 249(4):573--600, 2020.

\bibitem[Bur21]{Burylko2021}
O.~A. Burylko.
\newblock {Collective Dynamics and Bifurcations in Symmetric Networks of Phase
  Oscillators. II}.
\newblock {\em Journal of Mathematical Sciences (United States)},
  253(2):204--229, 2021.

\bibitem[BYMS21]{BerYanMaiSch2021}
Rico Berner, Serhiy Yanchuk, Yuri Maistrenko, and Eckehard Sch{\"o}ll.
\newblock Generalized splay states in phase oscillator networks.
\newblock {\em Chaos: An Interdisciplinary Journal of Nonlinear Science},
  31(7):073128, 2021.

\bibitem[CG88]{ChGo1988}
P.~Chossat and M.~Golubitsky.
\newblock Symmetry-increasing bifurcation of chaotic attractors.
\newblock {\em Physica D: Nonlinear Phenomena}, 32(3):423--436, 1988.

\bibitem[CP20]{CluPo2020}
Pau Clusella and Antonio Politi.
\newblock {Irregular collective dynamics in a Kuramoto--Daido system}.
\newblock {\em Journal of Physics: Complexity}, 2(1):014002, 2020.

\bibitem[CPR16]{CPR16}
Pau Clusella, Antonio Politi, and Michael Rosenblum.
\newblock A minimal model of self-consistent partial synchrony.
\newblock {\em New Journal of Physics}, 18(9):093037, sep 2016.

\bibitem[Dat18]{Datseris2018}
George Datseris.
\newblock {DynamicalSystems.jl: A Julia software library for chaos and
  nonlinear dynamics}.
\newblock {\em Journal of Open Source Software}, 3(23):598, mar 2018.

\bibitem[DGK{\etalchar{+}}08]{DhGoKuMeiSau}
A.~Dhooge, W.~Govaerts, Yu.~A. Kuznetsov, H.~G.E. Meijer, and B.~Sautois.
\newblock {New features of the software MatCont for bifurcation analysis of
  dynamical systems}.
\newblock {\em Mathematical and Computer Modelling of Dynamical Systems},
  14(2):147--175, 2008.

\bibitem[EM14]{EngelMir2014}
Jan~R. Engelbrecht and Renato Mirollo.
\newblock {Classification of attractors for systems of identical coupled
  Kuramoto oscillators}.
\newblock {\em Chaos: An Interdisciplinary Journal of Nonlinear Science},
  24(1):013114, 2014.

\bibitem[ESF18]{endres2018simplicial}
Stefan Endres, Carl Sandrock, and Walter Focke.
\newblock {A simplicial homology algorithm for Lipschitz optimisation}.
\newblock {\em Journal of Global Optimization}, 72(2):181--217, 2018.

\bibitem[GG16]{GonGon2016}
A.~S. Gonchenko and S.~V. Gonchenko.
\newblock {Variety of strange pseudohyperbolic attractors in three-dimensional
  generalized H{\'e}non maps}.
\newblock {\em Physica D: Nonlinear Phenomena}, 337:43--57, 2016.

\bibitem[GGKT14]{GonGonKazTur2014}
Alexander Gonchenko, Sergey Gonchenko, Alexey Kazakov, and Dmitry Turaev.
\newblock Simple scenarios of onset of chaos in three-dimensional maps.
\newblock {\em International Journal of Bifurcation and Chaos}, 24(08):1440005,
  2014.

\bibitem[GGS12]{GonGonShil2012}
A.~S. Gonchenko, S.~V. Gonchenko, and L.~P. Shilnikov.
\newblock Towards scenarios of chaos appearance in three-dimensional maps.
\newblock {\em Rus. Nonlin. Dyn.}, 8:3--28, 2012.

\bibitem[GKS22]{GrKazSat2022}
Evgeny~A Grines, Alexey Kazakov, and Igor~R Sataev.
\newblock {On the origin of chaotic attractors with two zero Lyapunov exponents
  in a system of five biharmonically coupled phase oscillators}.
\newblock {\em Chaos: An Interdisciplinary Journal of Nonlinear Science},
  32(9):093105, 2022.

\bibitem[GO18]{GO18}
Evgeny~A. Grines and Grigory~V. Osipov.
\newblock {Heteroclinic and Homoclinic Structures in the System of Four
  Identical Globally Coupled Phase Oscillators with Nonpairwise Interactions}.
\newblock {\em Regular and Chaotic Dynamics}, 23(7-8):974–982, Dec 2018.

\bibitem[GTRP20]{GenTeiRosPik2020}
Erik Gengel, Erik Teichmann, Michael Rosenblum, and Arkady Pikovsky.
\newblock High-order phase reduction for coupled oscillators.
\newblock {\em Journal of Physics: Complexity}, 2(1):015005, nov 2020.

\bibitem[GV04]{GuckVlad2004}
John Guckenheimer and Alexander Vladimirsky.
\newblock {A Fast Method for Approximating Invariant Manifolds}.
\newblock {\em SIAM Journal on Applied Dynamical Systems}, 3(3):232--260, jan
  2004.

\bibitem[Hau21]{Haug2021}
Sindre~W Haugland.
\newblock The changing notion of chimera states, a critical review.
\newblock {\em Journal of Physics: Complexity}, 2(3):032001, 2021.

\bibitem[HMM93]{HMM93}
D.~Hansel, G.~Mato, and C.~Meunier.
\newblock Clustering and slow switching in globally coupled phase oscillators.
\newblock {\em Physical Review E}, 48(5):3470--3477, November 1993.

\bibitem[KB02]{KurBat2002}
Y~Kuramoto and D~Battogtokh.
\newblock Coexistence of coherence and incoherence in nonlocally coupled phase
  oscillators.
\newblock {\em Nonlinear Phenomena in Complex Systems}, 5(4):380--385, 2002.

\bibitem[KK01]{KK01}
Hiroshi Kori and Yoshiki Kuramoto.
\newblock Slow switching in globally coupled oscillators: robustness and
  occurrence through delayed coupling.
\newblock {\em Phys. Rev. E}, 63:046214, Mar 2001.

\bibitem[KPS{\etalchar{+}}22]{KoPaSeTrVr2022}
Ilias~S. Kotsireas, Panos~M. Pardalos, Alexander Semenov, William~T. Trevena,
  and Michael~N. Vrahatis.
\newblock {Survey of Methods for Solving Systems of Nonlinear Equations, Part
  II: Optimization Based Approaches}, 2022.

\bibitem[Kur75]{Kur1975}
Yoshiki Kuramoto.
\newblock Self-entrainment of a population of coupled non-linear oscillators.
\newblock In {\em International symposium on mathematical problems in
  theoretical physics}, pages 420--422. Springer, 1975.

\bibitem[Kur84]{kuramoto1984chemical}
Yoshiki Kuramoto.
\newblock Chemical turbulence.
\newblock In {\em Chemical oscillations, waves, and turbulence}, pages
  111--140. Springer, 1984.

\bibitem[LP19]{LeonPazo2019}
Iv\'an Le\'on and Diego Paz\'o.
\newblock {Phase reduction beyond the first order: The case of the mean-field
  complex Ginzburg-Landau equation}.
\newblock {\em Phys. Rev. E}, 100:012211, Jul 2019.

\bibitem[LP22]{LeonPazo2022}
Iv\'an Le\'on and Diego Paz\'o.
\newblock {Enlarged Kuramoto model: Secondary instability and transition to
  collective chaos}.
\newblock {\em Phys. Rev. E}, 105:L042201, Apr 2022.

\bibitem[MPG22]{MajPercGhosh2022}
Soumen Majhi, Matja{\v{z}} Perc, and Dibakar Ghosh.
\newblock {Dynamics on higher-order networks: a review}.
\newblock {\em Journal of the Royal Society Interface}, 19(188):20220043, 2022.

\bibitem[Ome18]{Omel2018}
O~E Omel’chenko.
\newblock The mathematics behind chimera states.
\newblock {\em Nonlinearity}, 31(5):R121, apr 2018.

\bibitem[PJA{\etalchar{+}}21]{Parastesh2021}
Fatemeh Parastesh, Sajad Jafari, Hamed Azarnoush, Zahra Shahriari, Zhen Wang,
  Stefano Boccaletti, and Matja{\v{z}} Perc.
\newblock {Chimeras}.
\newblock {\em Physics Reports}, 898:1--114, 2021.

\bibitem[PMT05]{PopMaistrTass2005}
Oleksandr~V. Popovych, Yuri~L. Maistrenko, and Peter~A. Tass.
\newblock Phase chaos in coupled oscillators.
\newblock {\em Phys. Rev. E}, 71:065201, Jun 2005.

\bibitem[PR15]{PikRos2015}
Arkady Pikovsky and Michael Rosenblum.
\newblock {Dynamics of globally coupled oscillators: Progress and
  perspectives}.
\newblock {\em Chaos: An Interdisciplinary Journal of Nonlinear Science},
  25(9):097616, 2015.

\bibitem[SK86]{KurSak1986}
Hidetsugu Sakaguchi and Yoshiki Kuramoto.
\newblock {A soluble active rotater model showing phase transitions via mutual
  entertainment}.
\newblock {\em Progress of Theoretical Physics}, 76(3):576--581, 1986.

\bibitem[SO15]{SudaOkuda2015}
Yusuke Suda and Koji Okuda.
\newblock Persistent chimera states in nonlocally coupled phase oscillators.
\newblock {\em Physical review E}, 92(6):060901, 2015.

\bibitem[SPMS17]{StankPerMcClintStefa2017}
Tomislav Stankovski, Tiago Pereira, Peter V.~E. McClintock, and Aneta
  Stefanovska.
\newblock {Coupling functions: Universal insights into dynamical interaction
  mechanisms}.
\newblock {\em Rev. Mod. Phys.}, 89:045001, Nov 2017.

\bibitem[TP02]{TopPik2002}
Dmitri Topaj and Arkady Pikovsky.
\newblock {Reversibility vs. synchronization in oscillator lattices}.
\newblock {\em Physica D: Nonlinear Phenomena}, 170(2):118--130, 2002.

\bibitem[Tre84]{Tresser1984}
C.~Tresser.
\newblock {About some theorems by L. P. Šil'nikov}.
\newblock {\em Annales de l'I.H.P. Physique théorique}, 40(4):441--461, 1984.

\bibitem[Win67]{Winfree1967}
Arthur~T. Winfree.
\newblock Biological rhythms and the behavior of populations of coupled
  oscillators.
\newblock {\em Journal of Theoretical Biology}, 16(1):15 -- 42, 1967.

\bibitem[Win80]{winfree1980}
Arthur~T Winfree.
\newblock {\em The geometry of biological time}, volume~2.
\newblock Springer, 1980.

\bibitem[WO11]{WolfOmel2011}
Matthias Wolfrum and Oleh~E. Omel'chenko.
\newblock Chimera states are chaotic transients.
\newblock {\em Phys. Rev. E}, 84:015201, Jul 2011.

\bibitem[WS93]{WatStro1993}
S.~Watanabe and S.~Strogatz.
\newblock Integrability of a globally coupled oscillator array.
\newblock {\em Phys. Rev. Lett.}, 70:2391--2394, Apr 1993.

\end{thebibliography}

\newpage
\section*{Appendix. The algorithm for searching approximate heteroclinic cycles.}
\label{sec:appendix}


At the first step, we find all the equilibrium states lying on the faces of the invariant tetrahedron $\mathcal{C}$. Since the faces of the tetrahedron are mapped into each other under the action of the symmetry $T$ and its powers, it suffices to find the equilibrium states only on one of them. Without loss of generality, let us consider the face $\psi_2 = 0$.

Using the notation of formula \eqref{eq:RedusedA4d}, we write the system of differential equations corresponding to the restriction of system \eqref{eq:RedusedA4d} to the invariant plane $\psi_2 = 0$ as
\begin{equation}
\label{eq:PlaneRestriction}
\begin{cases}
\dot{\psi_3} = \Psi_3(0, \psi_3, \psi_4) = P(\psi_3,\psi_4)\\
\dot{\psi_4} = \Psi_4(0, \psi_3, \psi_4) = Q(\psi_3,\psi_4).
\end{cases}
\end{equation}
To find the coordinates of equilibrium states, we use the global optimization method called SHGO (Simplicial Homology Global Optimization) \cite{endres2018simplicial} and apply it to the objective function
\begin{equation}
\label{eq:ScalarFunc}
 \mathcal{F} (\psi_3, \psi_4) = P^2(\psi_3,\psi_4) + Q^2(\psi_3,\psi_4).
\end{equation}
The points of global minima at which this function equals zero correspond to the equilibrium states (see, for example, \cite{KoPaSeTrVr2022}) of system \eqref{eq:PlaneRestriction}.
This correspondence is exact from a mathematical point of view, however, due to the finite accuracy of floating-point arithmetic, we consider a point to be the equilibrium state of system \eqref{eq:PlaneRestriction} if the objective function \eqref{eq:ScalarFunc} takes in it a value that is less than some preassigned $\epsilon$. Next, we check whether there are saddles and saddle-foci with an appropriate configuration and dimensions of stable and unstable manifolds among these equilibria.
If such saddles and saddle-foci are not found, then we stop the analysis, otherwise we proceed to the second step.

At the second step, we iterate over all possible pairs of saddles and saddle-foci and calculate the value of $\tilde{p}$ for each pair using formula \eqref{eq:ptild}.
By Tresser pairs we will call such pairs that satisfy $\tilde{p} > 1$ condition.
Let us designate the equilibrium states in this pair as $\widehat{\sigma}_{\rm s}$ (saddle) and $\widehat{\sigma}_{\rm sf}$ (saddle-focus).
As mentioned earlier, the value $\tilde{p}$ characterizes whether a complex dynamics is possible in the vicinity of the heteroclinic cycle.
Note that at this stage we know nothing about the existence of a heteroclinic cycle based on the equilibrium states $\widehat{\sigma}_{\rm s}$,\; $\widehat{\sigma}_{\rm sf}$ and their symmetrical copies.
However, calculating the value of $\tilde{p}$ before the costly (by the standards of the estimated time) verification of the existence of (approximate) heteroclinic connections allows us to answer the following question: if a heteroclinic cycle based on this Tresser pair exists, would that mean the presence of complex dynamics in the system?
Since in this paper the emphasis is put on the search for chaos based on the conditions of Tresser's theorem, only the variant $\tilde{p} > 1$ is of interest to us.
Such an order of checks can significantly save calculation time.
If no Tresser pairs were found at this step, then the analysis of the system is stopped, otherwise we proceed to the third step.

At the third step, for each of the Tresser pairs we finally analyze the behavior of one-dimensional separatrices.
First, we check whether the restriction of the system to the invariant face contains a separatrix connecting the equilibrium $\widehat{\sigma}_{\rm s}$ with the equilibrium $\widehat{\sigma}_{\rm sf}$.
To approximate the separatrix, we calculate the unstable eigenvector $\vec{v}_{\rm unst}$ of the saddle $\widehat{\sigma}_{\rm s}$ and assume that the points $\widehat{\sigma}_{\rm s} \, \pm \, \varepsilon \,  \vec{v}_{\rm unst}$ belong to separatrices for some small $\varepsilon$ (see, for example, \cite{GuckVlad2004}).
When the system is restricted to an invariant face, the saddle-focus $\widehat{\sigma}_{\rm sf}$ becomes just a stable focus, hence it suffices to calculate the separatrices and check whether they reach some small neighborhood of the stable focus.
If none of the separatrices has reached the given focus, then we discard this Tresser pair: there is no heteroclinic connection between the saddle and the saddle-focus, which means that they are not part of the same heteroclinic cycle.
If one of the separatrices has reached a given focus, then we proceed to the analysis of the behavior of the unstable separatrix of the saddle-focus $\widehat{\sigma}_{\rm sf}$.
Using the unstable eigenvector, we determine which of the saddle-focus separatrices remains in the canonical invariant region $\mathcal{C}$ and calculate its approximation according to the principle described earlier.
In order to avoid the difficulties associated with the numerical integration of the system of differential equations in the vicinity of known saddle equilibrium states, we stop the calculation when the trajectory enters some programmatically specified neighborhoods of these equilibria.
If the trajectory has entered the neighborhood of the saddle $\widehat{\sigma}_{\rm s}$ or some of its symmetric copies $T^{k}(\widehat{\sigma}_{\rm s})$, then we claim that we have found an approximate heteroclinic trajectory, otherwise we exclude this Tresser pair from consideration.
In case of success at each of the three steps, we get a set of equilibrium states and connections between them $\widehat{\sigma}_{\rm s} \rightarrow \widehat{\sigma}_{\rm sf} \rightarrow T^k(\widehat{\sigma}_{\rm s})$.
As we pointed out earlier, the presence of these two heteroclinic trajectories is sufficient to assert the existence of an approximate heteroclinic cycle.
Moreover, for this heteroclinic cycle, the value $p > 1$ (since $\tilde{p} > 1$), that is, in its vicinity, the existence of a countable number of closed trajectories is possible, which leads to the complex dynamics in the system.
Due to the finite precision of floating-point arithmetic, we cannot assert that a real (not approximate) heteroclinic cycle exists precisely for these parameter values.
However, the parameter values at which Tresser pairs with suitable heteroclinic connections were found could be used to localize regions where true heteroclinic cycles exist, and chaotic dynamics is also possible.
\end{document}